\documentclass[twocolumn,prb,superscriptaddress,floatfix]{revtex4-2}
\usepackage{amsmath,amssymb}
\usepackage{graphicx}
\usepackage{bm}
\usepackage{comment}
\usepackage{cancel}
\usepackage{xcolor}
\usepackage[colorlinks=true,citecolor=blue,urlcolor=blue]{hyperref}

\begin{document}

\title{Low-temperature spin dynamics in LAFO thin films: from cubic anisotropy to TLS-limited coherence}

\author{Srishti Pal}
\thanks{These authors contributed equally to this work.}
\affiliation{School of Applied and Engineering Physics, Cornell University, Ithaca, NY 14853, USA}

\author{Guanxiong Qu}
\thanks{These authors contributed equally to this work.}
\affiliation{Department of Physics and Astronomy, University of California, Irvine, CA 92697, USA}

\author{Herv\'{e} M. Carruzzo}
\thanks{These authors contributed equally to this work.}
\affiliation{Department of Physics and Astronomy, University of California, Irvine, CA 92697, USA}

\author{Katya Mikhailova}
\affiliation{Department of Applied Physics, Stanford University, Stanford, California 94305, USA}
\affiliation{Geballe Laboratory for Advanced Materials, Stanford University, Stanford, California 94305, USA}
\affiliation{Stanford Institute for Materials and Energy Sciences, SLAC National Accelerator Laboratory, 2575 Sand Hill Road, Menlo Park, California, 94025, USA}

\author{Lerato Takana}
\affiliation{Department of Applied Physics, Stanford University, Stanford, California 94305, USA}
\affiliation{Geballe Laboratory for Advanced Materials, Stanford University, Stanford, California 94305, USA}

\author{Qin Xu}
\affiliation{Department of Physics, Cornell University, Ithaca, NY 14853, USA}

\author{Yuri Suzuki}
\affiliation{Department of Applied Physics, Stanford University, Stanford, California 94305, USA}
\affiliation{Geballe Laboratory for Advanced Materials, Stanford University, Stanford, California 94305, USA}
\affiliation{Stanford Institute for Materials and Energy Sciences, SLAC National Accelerator Laboratory, 2575 Sand Hill Road, Menlo Park, California, 94025, USA}

\author{Clare C. Yu}
\thanks{Corresponding authors: cyu@uci.edu, gdf9@cornell.edu}
\affiliation{Department of Physics and Astronomy, University of California, Irvine, CA 92697, USA}

\author{Gregory D. Fuchs}
\thanks{Corresponding authors: cyu@uci.edu, gdf9@cornell.edu}
\affiliation{School of Applied and Engineering Physics, Cornell University, Ithaca, NY 14853, USA}
\affiliation{Kavli Institute at Cornell for Nanoscale Science, Cornell University, Ithaca, NY 14853, USA}

\date{\today}

\begin{abstract}

We investigate the low-temperature spin dynamics of epitaxial lithium aluminum ferrite (LAFO) thin films using broadband ferromagnetic resonance (FMR) spectroscopy from 0.44 K to 68 K. The results reveal a crossover from conventional cubic anisotropy-dominated behavior at higher temperatures to pronounced linewidth broadening and higher-order anisotropy contributions at cryogenic temperatures. With the magnetic field oriented along the [100] crystallographic direction, the resonance is well-captured by four-fold in-plane and out-of-plane uniaxial anisotropies. In contrast, measurements with the field along the [110] direction reveal the presence of an unusually large sixth-order cubic anisotropy term that is symmetry-suppressed for [100] but becomes apparent under this field orientation at ultralow temperatures, indicating a substantial modification of the anisotropy landscape. Independent linewidth analysis shows a pronounced peak near 8 K and a subtle monotonic enhancement with decreasing temperatures below 2 K, features consistent with dissipation mediated by a bath of two-level systems (TLS) arising from antisite defects and localized Fe$^{3+}$ moments. Comparison with TLS-based models demonstrates that both exchange-coupled impurities and nearly free paramagnetic centers contribute to the observed damping. Our results establish LAFO as a model ferrite system where disorder-induced TLS limit spin coherence at ultralow temperatures, providing new insights into anisotropy engineering, magnetic relaxation, and the design of ferrimagnetic insulators for coherent magnonics. These findings offer a framework for future optimization of growth conditions.

\end{abstract}

\maketitle

\section{Introduction}
\label{MS:Intro}

Magnetically ordered insulators have emerged as a fertile platform for exploring coherent spin dynamics and magnon-based information processing. In these systems, spin excitations propagate without charge current, eliminating Joule heating and enabling low-dissipation transport of information. Such properties make them attractive for a broad range of technologies from microwave signal processing to hybrid quantum devices~\cite{Brataas2020,Pirro2021,Cornelissen2015}. Central to these applications is the ferromagnetic resonance (FMR) linewidth, which is related to magnetic damping and ultimately sets the coherence time of magnons~\cite{Pfirrmann2019,Kosen2019,Hong2024}. Identifying mechanisms that contribute to damping and discovering routes to suppress them remain key challenges in advancing ferrimagnetic insulators toward practical and quantum-coherent regimes.  

Among magnetic insulators, garnet ferrites such as yttrium iron garnet (YIG) and spinel ferrites, such as copper ferrite (CuFe$_2$O$_4$) and lithium ferrite (LiFe$_2$O$_4$, LFO), have long been recognized for their exceptionally low damping, narrow linewidths, and tunable magnetic anisotropies~\cite{Hauser2016,Dubs2020,Pachauri2015,Zheng2020}. LFO in particular offers several interesting properties: it is a ferrimagnetic oxide with a relatively simple crystal structure, high Curie temperature, and compatibility with thin-film growth on lattice-matched substrates. Moreover, chemical substitution provides a route to engineer magnetic anisotropy and the mechanisms governing how the magnetization of the host lattice spins relax. A prominent example is aluminum-substituted lithium ferrite (LAFO), in which Al$^{3+}$ ions preferentially occupy octahedral sites, modifying exchange interactions, magnetocrystalline anisotropy, and damping~\cite{Zheng2023,OMahoney2023}. However, substitution can also promote antisite disorder in which ions or atoms of two different elements switch lattice sites. This is a well-documented feature of ferrites~\cite{Gilleo1958}, particularly in spinel systems, arising when cations partially exchange their equilibrium sites within the tetrahedral and octahedral sublattices~\cite{Goya1998,Goldman2006,Askarzadeh2024}. Such disorder can place Fe$^{3+}$ ions at non-equilibrium positions, giving rise to localized magnetic moments whose behavior depends on their immediate atomic environment — namely, the nature of neighboring ions, bond geometry, and the resulting exchange pathways. When the local coordination allows sufficient orbital overlap with nearby Fe$^{3+}$ ions, these antisite moments can remain exchange-coupled to the ferrimagnetic lattice~\cite{Matsubara2020}. In contrast, if the surrounding sites are predominantly nonmagnetic or the exchange pathways are disrupted by structural distortion, the Fe$^{3+}$ spins become weakly interacting, behaving as paramagnetic impurities~\cite{Werner2020}. Both types of impurities have the potential to degrade spin coherence at low temperatures.

While the room-temperature FMR response of ferrites is relatively well characterized, their behavior at cryogenic temperatures remains less well understood. In the conventional Gilbert damping picture, dissipation is described solely by viscous damping and static inhomogeneous broadening~\cite{Flaig2017}. Recent studies have highlighted deviations from this picture under certain conditions, including temperature-dependent linewidths, nonmonotonic evolution, and signatures of interaction with extrinsic baths~\cite{Pfirrmann2019,Jermain2017}. One prominent mechanism is coupling to parasitic two-level systems (TLS). The concept of TLS, originally developed in the late 1960s and early 1970s to explain anomalous low-temperature thermal and acoustic properties of amorphous solids~\cite{Zeller1971,Anderson1972,Hunklinger1986,phillips1987}, has since been extended to crystalline materials. These TLS arise from atomic tunneling centers, structural defects, or localized impurity spins, and are now recognized as a ubiquitous source of decoherence in condensed-matter systems~\cite{Martinis2005,Lisenfeld2015,Muller2019}. In superconducting qubits and microwave resonators, they are known to dominate energy relaxation and noise at millikelvin temperatures~\cite{Martinis2005,Muller2019,Gao2008}. Their manifestation in magnetic insulators is still being be explored. However, growing evidence suggests that TLS can couple to magnon modes, providing an additional, temperature-dependent damping channel~\cite{Pfirrmann2019,Kosen2019,Tabuchi2014}.  

In ferrites, antisite disorder and local lattice distortions naturally produce defect spins that may act as TLS. For instance, CoFe$_2$O$_4$ films show modified anisotropy due to antisite disorder~\cite{Omelyanchik2020}, while in LiFePO$_4$, Fe ions occupying Li sites contribute quasi-free local moments that broaden the linewidth~\cite{Werner2020}. These studies highlight that localized spins in disordered environments can mimic TLS-like dynamics, thereby influencing both relaxation and anisotropy in complex ways. Yet, a systematic examination of TLS-mediated damping in ferrite thin films, particularly at cryogenic temperatures, remains lacking.  

In this work, we investigate the low-temperature ferromagnetic resonance of epitaxial 15 nm LAFO thin films grown on MgAl$_2$O$_4$ (MAO) substrates down to 0.44 K. By tracking resonance conditions and linewidths with the magnetic field oriented along the [100] and [110] crystallographic directions, we uncover three central findings. First, the [100] resonance is well described by conventional four-fold cubic and uniaxial anisotropy terms. In contrast, measurements with the field applied along the [110] direction reveal a pronounced sixth-order cubic anisotropy contribution that is symmetry-suppressed for [100]. The enhanced visibility of this higher-order term along [110] points to the sensitivity of this orientation to disorder-induced local symmetry perturbations, which lift cancellations present along high-symmetry directions and amplify higher-order anisotropic contributions at low temperatures. Second, the linewidth exhibits a pronounced peak near 8 K and a slight upturn below 2 K, inconsistent with purely intrinsic damping but naturally explained by TLS-induced relaxation. Third, by correlating anisotropy evolution with linewidth behavior, we identify contributions from both exchange-coupled defect spins and nearly isolated paramagnetic centers, each dominating different temperature regimes.  

Our results establish LAFO as a model ferrite system in which defect-induced TLS set the fundamental limit of magnon coherence at low temperatures. Beyond clarifying the role of disorder and high-order anisotropy in ferrites, this work provides new insights into the microscopic origins of damping in insulating magnets. More broadly, our findings have implications for the design of low-loss magnetic insulators for coherent magnonics, quantum transduction, and hybrid spin–photon platforms.

\section{Experimental Methods}
\label{MS:Exp Methods}

Epitaxial thin films of LAFO were synthesized on (001)-oriented single-crystalline MgAl$_2$O$_4$ substrates using pulsed laser deposition with a 248 nm KrF laser at a laser fluence of 2.8 J/cm$^2$. The substrates were ultrasonically cleaned in acetone and isopropanol for 5 minutes each. A Toshiba cooperation polycrystalline target of Li$_{0.6}$Al$_{1.0}$Fe$_{1.5}$O$_{4.0}$ was sanded and pre-ablated, for 1 minute at 1 Hz plus 1.5 minutes at 5 Hz, before growth. Excess Li is used to account for Li volatility at higher temperatures. The growth was performed at a substrate temperature of 425$^{\circ}$C in 15 mTorr of oxygen followed by a cool down to rrom temperature in 150 Torr of oxygen. Temperature and field dependent magnetization measurements of the grown film were performed on a SQUID magnetometer and are shown in Appendix~\ref{Appendix:Mat Char}.

Broadband ferromagnetic resonance (FMR) measurements were performed using a Keysight E5063A vector network analyzer (VNA) connected to a broadband coplanar waveguide fabricated from Ti/Cu/Pt multilayers (fabrication details provided in Appendix~\ref{Appendix:Waveguide Fab}). The film was mounted face down on the waveguide using cryogenic Apiezon N grease to ensure good thermal contact and stable mechanical alignment. The external magnetic field was applied in the film plane, with the crystallographic orientation controlled to align either along the [100] or [110] axis. The magnetic field was calibrated following the procedure described in Ref.~\cite{Qu2024}.

All measurements were carried out in a Bluefors SD $^3$He cryostat. The base temperature of 0.44 K was achieved under active $^3$He circulation. The sample temperature was varied from 0.44 K to 1.35 K by applying controlled heating to the still stage while maintaining the $^3$He pressure below 1 atm. Above this range, the $^3$He circulation was stopped, and the system operated in $^4$He mode, allowing measurements between 3.1 K and 68 K. The upper temperature limit was set by the requirement to keep the superconducting electromagnet below its critical temperature of 4.2 K.

\section{Results}
\label{MS:Results}

\subsection{Representative FMR spectra}
\label{MS:Results-Raw FMR}

\begin{figure}[h!]
\centering
\includegraphics[width=\columnwidth]{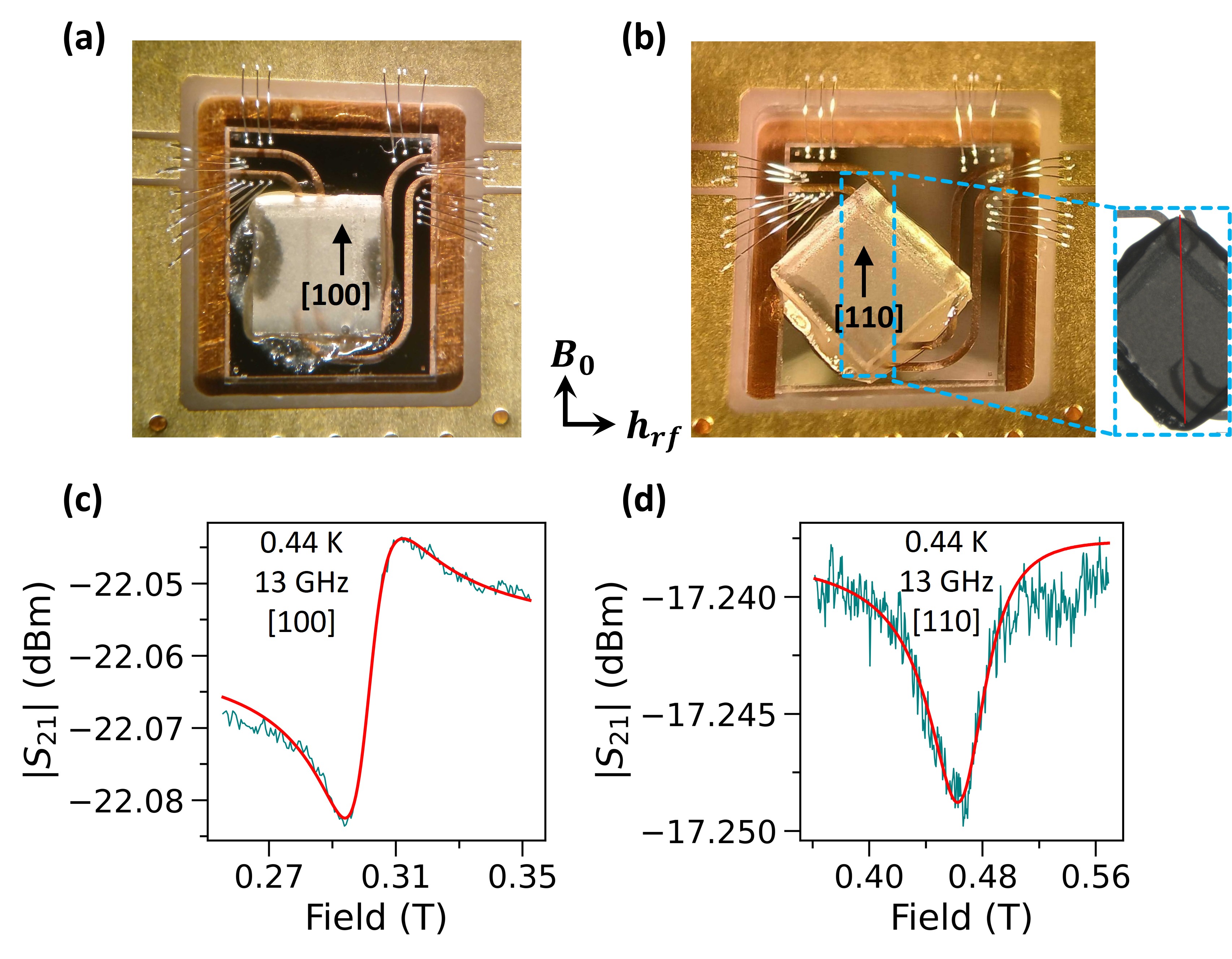}
\caption{(a)-(b) Microscope images of the 15 nm LAFO/MAO flipped on broadband coplanar waveguide chip along [100] and [110] crystallographic directions, respectively. The magnified microscope image in (b) shows the alignment of the sample along the [110] direction. (c)-(d) Broadband FMR response (dark cyan curve) at 13 GHz at 0.44 K fit to a combination of absorptive and dispersive contributions (red solid curves) along [100] and [110] directions, respectively.} 
\label{fig:fig1}
\end{figure}

We first establish the quality and symmetry dependence of the low-temperature FMR response by comparing representative spectra measured with the field along the [100] and [110] crystallographic directions at base temperature. These spectra demonstrate well-defined resonances with distinct linewidths and lineshapes that reflect the underlying anisotropy and disorder-sensitive magnetic response of the film. Figure~\ref{fig:fig1} shows microscope images of the 15\,nm LAFO/MAO film flipped on the coplanar waveguide and representative broadband FMR spectra at 13\,GHz at 0.44\,K for fields along [100] and [110]. The resonances were fitted using a Lorentzian model that includes absorptive and dispersive parts~\cite{Yadav2014}
\begin{equation}
\begin{aligned}
S_{21}(H)
&=
L\,\frac{\Delta H_{\mathrm{FWHM}}^{2}}
{(H-H_{\mathrm{fmr}})^{2}+\Delta H_{\mathrm{FWHM}}^{2}}
\\
&\quad
+\,D\,\frac{\Delta H_{\mathrm{FWHM}}(H-H_{\mathrm{fmr}})}
{(H-H_{\mathrm{fmr}})^{2}+\Delta H_{\mathrm{FWHM}}^{2}}
+ C,
\end{aligned}
\label{eq:FMR_Fit}
\end{equation}

\noindent{where $L$ and $D$ are the amplitudes of the absorptive and dispersive components, $H_{fmr}$ is the resonance field, and $\Delta H_{FWHM}$ is the full width at half maximum.}

\subsection{Ferromagnetic resonance frequency: anisotropy fits}
\label{MS:Results-Aniso Fits}

\begin{figure}[h!]
\centering
\includegraphics[width=\columnwidth]{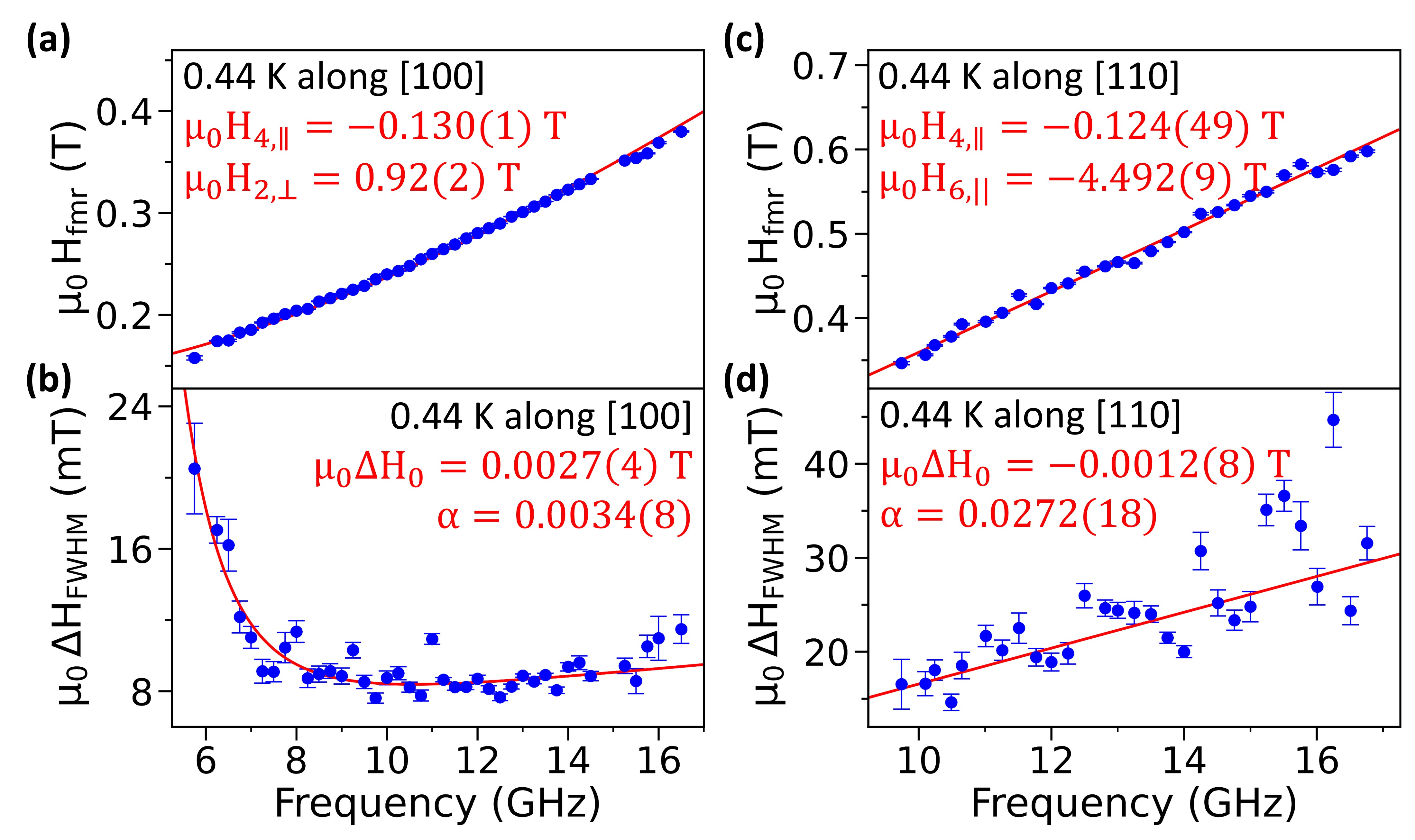}
\caption{(a), (c) Frequency dependence of resonance field at 0.44 K with fits to the Smit-Beljers forms for [100] and [110] directions, respectively. (b), (d) Fitted frequency variation of FMR linewidth at 0.44 K along [100] and [110] directions, respectively. The error bars shown correspond to the standard errors obtained from the nonlinear least-squares fits of the FMR spectrum using Eq.~\ref{eq:FMR_Fit} at each frequency.}
\label{fig:fig2}
\end{figure}

We next analyze the resonance frequency as a function of applied magnetic field to extract the magnetic anisotropy parameters governing the spin dynamics at cryogenic temperatures. The frequency-field relationship with the field along [100] and [110] as shown in Fig.~\ref{fig:fig2}(a) and (c) is fit using a Smit–Beljers expression~\cite{Smit1955} that includes fourth and sixth order cubic anisotropies in the plane ($H_{4,\parallel}$ and $H_{6,\parallel}$) and a uniaxial anisotropy out of the plane $H_{2,\perp}$. Under the condition that the external field $H$ is strong enough to saturate the magnetization, the resonance frequencies along the [100] and [110] directions can be written as
\begin{equation}
\begin{aligned}
\omega_{0}
&= \gamma \mu_{0}
\sqrt{
H + H_{4,\parallel}\cos(4\phi_{H})
}
\\
&\!\times
\sqrt{
H + M_{\mathrm{eff}}
+ \frac{1}{4} H_{4,\parallel}\left(3+\cos 4\phi_{H}\right)
+ \frac{1}{4} H_{6,\parallel}\sin^{2}(2\phi_{H}),
}
\end{aligned}
\label{eq:SB_Fit}
\end{equation}

\noindent{where $\gamma$ is the gyromagnetic ratio and $M_{\mathrm{eff}} = M_s + \frac{2K_u}{\mu_0M_S} = M_s + H_{2,\perp}$. $K_u$ is the uniaxial magnetic anisotropy constant. $\phi_H$ is the angle of the external magnetic field with respect to the [100] crystallographic direction. While $H_{6,\parallel}$ is non-zero along [110] $(\phi_H = \pi/4)$, its contribution to the FMR frequency vanishes along [100] $(\phi_H = 0)$.}

Fits to the [100] data at 0.44 K (red solid curve in Fig.~\ref{fig:fig2}(a)) are well described by the above form with a modest in-plane cubic anisotropy $\mu_0H_{4,||}$ of -0.130(1) T and a uniaxial anisotropy $\mu_0H_{2,\perp}$ of 0.92(2) T (using a saturation magnetization $\mu_0M_S$ of 0.13 T from the SQUID data along [100] shown in Appendix~\ref{Appendix:Mat Char}). Notably, the anisotropy parameters extracted for the [100] direction at 0.44 K fall within a physically reasonable range relative to the room-temperature values (Appendix~\ref{Appendix:RT FMR}).

The frequency dependence of the [110] data, however, exhibits a pronounced offset relative to the [100] data at 0.44 K, in contrast to the room-temperature behavior shown in Appendix~\ref{Appendix:RT FMR}. This behavior indicates the presence of a sizable sixth-order anisotropic field $H_{6,\parallel}$ at cryogenic temperatures, which compensates the effective magnetization $M_{\mathrm{eff}}$ and results in an approximately linear frequency dependence, see Fig.~\ref{fig:fig2}(c). To fit the [110] data at 0.44 K, we fix $\mu_0M_{eff}$ to be 1.11 T (taking $\mu_0M_S$ of 0.19 T from the SQUID data along [110] shown in Appendix~\ref{Appendix:Mat Char} and $\mu_0H_{2,\perp}$ of 0.92 T extracted from fitting the [100] data). We fixed $H_{2,\perp}$ to the value from the [100] fit because it is expected to be identical for all in-plane orientations. Allowing it to vary in the [110] fit does not appreciably improve the fit quality but introduces large parameter covariance. The in-plane anisotropy coefficients are then extracted from the fit (red solid curve in Fig.~\ref{fig:fig2}(c)) as $\mu_0H_{4,||}$ of -0.124(49) T and $\mu_0H_{6,||}$ of -4.480(8) T. The estimated magnitude of the sixth-order anisotropy field turns out to be as large as $H_{6,\parallel} \sim -4M_{\mathrm{eff}}$. The minimal offset observed in the frequency dependence at room temperature (see Appendix~\ref{Appendix:RT FMR}) suggests that the sixth-order anisotropic field becomes quite small at higher temperatures as expected due to its $(M(T)/M(0))^{21}$ dependence~\cite{CALLEN1966}.

\subsection{Linewidth: frequency dependence}
\label{MS:Results-Linewidth Fits}

To disentangle intrinsic damping from extrinsic broadening mechanisms, we examine the frequency dependence of the FMR linewidth with the field along the crystallographic directions [100] and [110]. The full linewidth was modeled with the phenomenological form~\cite{OMahoney2023}
\begin{equation}
\begin{aligned}
\Delta H_{\mathrm{FWHM}}(f,T)
&=
\Delta H_{0}(T)
+ \alpha(T)\,
\frac{h}{g\,\mu_{B}\mu_{0}}\,f
\\
&\quad
+ \Delta H_{\mathrm{low}}(T)
\left(\frac{H_{z}}{f}\right)^{\beta},
\end{aligned}
\label{eq:linewidth_model}
\end{equation}

\noindent{where $\Delta H_0$ is the contribution from the inhomogeneous broadening and two-level impurity scattering, $\alpha$ is the effective (phenomenological) Gilbert damping parameter, and the low-field term $\Delta H_{\mathrm{low}}(H_z/f)^\beta$ represents the low-field loss due to incomplete saturation. The fitted frequency variation of the FMR linewidth for the field along [100] and [110] directions are shown in Fig.~\ref{fig:fig2}(b) and (d), respectively. Whereas for the [100] data, the low-field contribution is sizable at low frequencies, for the [110] data, the third term is negligible within the experimental uncertainty and a simple linear frequency dependence is observed.}

\subsection{Temperature dependence of fit parameters}
\label{MS:Results-Temperature Evolution}

\begin{figure}[h!]
\centering
\includegraphics[width=\columnwidth]{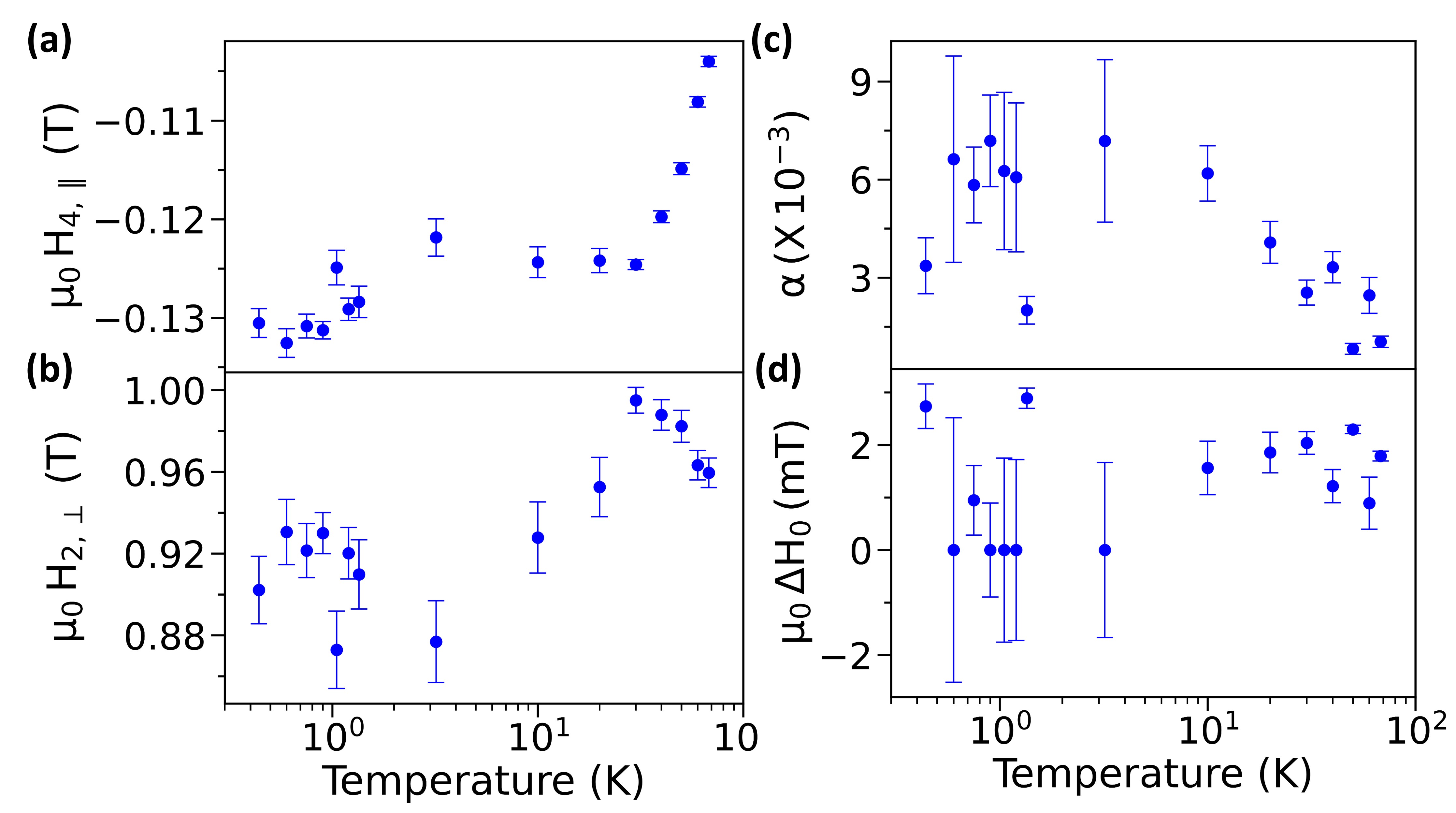}
\caption{(a)-(b) Temperature evolution of the anisotropy fields $H_{4,||}$ and $H_{2,\perp}$ measured with the magnetic field along [100]. (c)-(d) Temperature evolution of Gilbert damping $\alpha$ and the inhomogeneous broadening $\Delta H_0$ also with the field along [100]. Error bars indicate the standard errors from nonlinear least-squares fits performed for the resonance frequency and linewidth variations using Eqs.~\ref{eq:SB_Fit} and~\ref{eq:linewidth_model}, respectively.} 
\label{fig:fig3}
\end{figure}

We now track the temperature evolution of the extracted anisotropy and linewidth parameters to identify deviations from conventional behavior. The temperature evolution of the anisotropy fields $\mu_0H_{4,||}$ and $\mu_0H_{2,\perp}$, extracted from resonance frequency fits for the field applied along [100] from 0.44 K to 68 K, is shown in Fig.~\ref{fig:fig3} (a)-(b). Both in-plane and uniaxial anisotropies reveal non-monotonic behavior as a function of temperature, which are discussed in Sec.~\ref{MS:Discussion}.

In contrast to the [100] orientation, where measurements up to 68 K yielded a reliable temperature dependence, the [110] orientation could only provide data over a limited range of temperature. With increasing temperature, the signal weakened and broadened, leaving too few data points to extract meaningful fit parameters (see Appendix~\ref{Appendix:T-dep FMR [110]}).

Figure~\ref{fig:fig3} (c)-(d) shows the temperature dependence of $\alpha$ and $\Delta H_0$, extracted from the linewidth fits for the field applied along [100] from 0.44 K to 68 K. The inhomogeneous broadening, $\Delta H_0$, remains essentially temperature-independent with negligible values within the experimental uncertainty across the entire temperature range and limited to less than 30 G. In contrast, $\alpha$ along [100] exhibits a potential decrease at temperatures above $\sim$20 K. We note, however, that the quantitative extraction of $\alpha$ depends on the relative contributions of the low- and high-frequency data and is therefore somewhat unreliable given the restricted ($<$17 GHz) frequency range. On the other hand, the absolute linewidth above 12 GHz is reliable considering that the low frequency contributions are suppressed at $\sim$0.25~T and above.  In the further analysis we focus on the total linewidth for our analysis of temperature dependent trends.

\subsection{TLS simulation and comparison}
\label{MS:Results-TLS}

\begin{figure}[h!]
\centering
\includegraphics[width=\columnwidth]{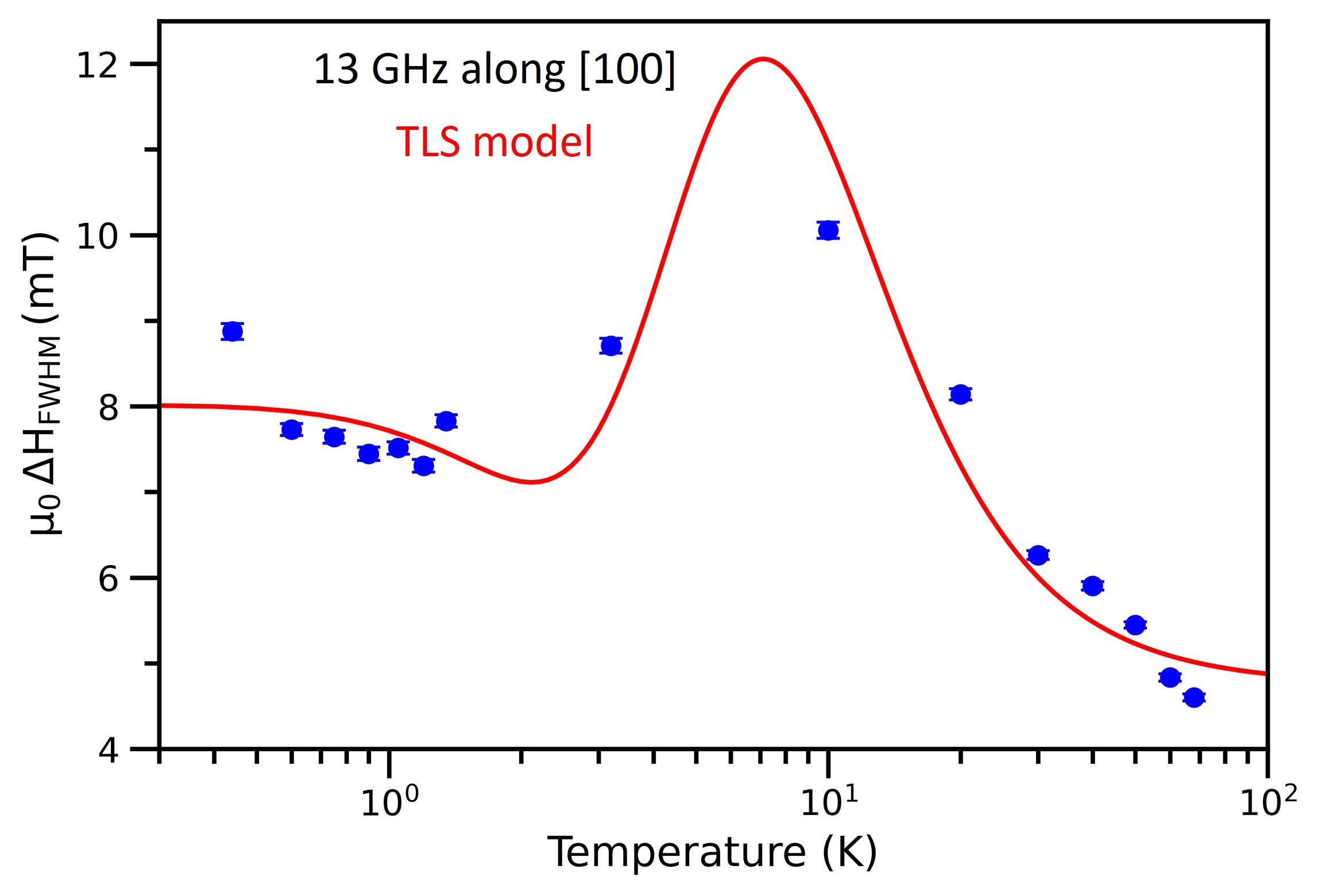}
\caption{TLS contribution to the temperature evolution of the linewidth with the field along [100] at 13 GHz. Blue symbols denote the experimentally obtained linewidths obtained from Lorentzian fits (Eq.~\ref{eq:FMR_Fit}) to the raw FMR spectra, with error bars representing the standard errors from the nonlinear least-square fittings. The red solid curve shows the result of fitting with a two-TLS model, as described in Section~\ref{MS:Results-TLS}.} 
\label{fig:fig4}
\end{figure}

In the following, we focus on the temperature dependence of the total FMR linewidth for the field along [100], which is obtained directly from spectral fits at each frequency and is most robust at higher frequencies where low-frequency loss contributions are negligible. In addition to the frequency dependence at a fixed temperature, the FMR linewidth exhibits a characteristic temperature evolution, showing a maximum around 8 K and a slight increase below approximately 2 K, as illustrated in Fig.~\ref{fig:fig4}. Such a peak-like temperature dependence can be attributed to the magnon dissipation caused by a bath of two-level systems (TLS), such as the magnetic impurities in rare-earth doped yttrium iron garnet compounds~\cite{Spencer1959,Seiden1964}. These TLS provide an additional contribution to the total linewidth~\cite{Pfirrmann2019, Tabuchi2014}, which exhibits strong temperature dependence due to the thermal occupation of the two-level states. TLS produce magnon dissipation via two different mechanisms: relaxation and resonance.  Relaxation arises when the oscillating magnetic field associated with the FMR modulates the energy splitting of the impurities and as a result, the energy level population of the impurities must readjust to the new equilibrium. This readjustment requires energy that is extracted from the magnons. At low temperatures when the impurities are largely in their ground state, the relaxation process is no longer a factor. As the temperature increases, the TLS approach saturation in which there is nearly equal populations of TLS in the ground and excited states. As a result, TLS population readjustment decreases along with magnon damping. In resonant scattering of the magnon, the magnon frequency matches the TLS energy splitting $\omega_{imp}$. As a result, a magnon is absorbed by the TLS which is excited from the ground state to the excited state. Which mechanism dominates depends on the energy splitting of the TLS compared to the FMR frequency, $\omega_0$. We show below that relaxation dominates at higher temperature while resonant processes dominate at temperatures below the maximum in the linewidth.

When resonant absorption dominates, the transverse components of the TLS spins are driven by the precession of the magnetization of the host lattice. The FMR linewidth arising from this resonant mechanism is given by (see Appendix~\ref{Appendix:TLS Model})
\begin{equation}
\begin{aligned}
\Delta H^{\mathrm{res}}_{\mathrm{TLS}}(T)
&=
\frac{c_{\mathrm{imp}}\,M_{0}}{\mu_{0}\gamma M_{0}}\,
\frac{J_{\perp}^{2}}{J_{\parallel}^{2}}\,
\frac{\omega_{\mathrm{ex}}^{2}}{\omega_{\mathrm{imp}}}
\\
&\quad\times
\frac{\omega_{0}\,\tau_{\mathrm{imp}}(T)}
{1+\bigl(\omega_{0}-\omega_{\mathrm{imp}}\bigr)^{2}
\tau_{\mathrm{imp}}^{2}(T)}
\,\tanh\!\left(
\frac{\hbar\,\omega_{\mathrm{imp}}}{2k_{B}T}
\right),
\end{aligned}
\label{eq:TLS-resonant}
\end{equation}

\noindent{where $c_{imp}$ is the concentration of TLS, $J_{\perp}$ ($J_{||}$) denote the transverse (longitudinal) components (with respect to the quantization axis of the host spins) of the exchange energy between TLS and the host lattice. $m_0$ and $M_0$ are the saturation magnetization of the TLS and host lattice, respectively. The energy splitting of the impurities, $\omega_{imp}$, is determined by both the external field and exchange interaction and can be approximated by $\omega_{imp}\approx \omega_{0}+ \omega_{ex}$, where $\omega_{ex}$ denotes the exchange interaction between magnetic impurities and host lattice.  $\tau_{imp} (T)=\tau_0 \tanh \left(\hbar\omega_{imp}/2 k_B T \right)$ is the relaxation time of the TLS. Note that the relaxation time of the TLS is temperature dependent, as the occupation of the two levels varies with temperature~\cite{phillips1987}.}

The relaxation process dominates when the TLS spins relax toward thermal equilibrium, while this equilibrium is continuously modulated by the molecular field of the host spins.
The FMR linewidth arising from this relaxation mechanism is given by (see Appendix~\ref{Appendix:TLS Model})
\begin{equation}
\begin{aligned}
\Delta H^{\mathrm{relax}}_{\mathrm{TLS}}(T)
&=
\frac{c_{\mathrm{imp}}\,M_{0}}{\mu_{0}\gamma M_{0}}\,
\frac{J_{\parallel}^{2}-J_{\perp}^{2}}{J_{\parallel}^{2}}\,
\omega_{\mathrm{ex}}
\\
&\quad\times
\frac{\omega_{0}\,\tau_{\mathrm{imp}}(T)}
{1+\omega_{0}^{2}\tau_{\mathrm{imp}}^{2}(T)}\,
\frac{\hbar\,\omega_{\mathrm{ex}}}{2k_{B}T}\,
\text{sech}^{2}\!\left(
\frac{\hbar\,\omega_{\mathrm{imp}}}{2k_{B}T}
\right).
\end{aligned}
\label{eq:TLS-relax}
\end{equation}

The relaxation mechanism requires anisotropic exchange between the TLS and the host lattice, \textit{i.e.}, $J_{||} \neq J_{\perp}$. Note that the resonant and relaxation modes can be distinguished by their distinct temperature dependences: the resonant contribution decreases monotonically with increasing temperature, whereas the relaxation mechanism exhibits a maximum at a characteristic temperature.

The TLS in the LAFO film can be identified as magnetic impurities induced by aluminum substitution, where the aluminum-surrounded environment isolates some Fe cations, thus modifying the exchange with the spins of the host lattice. The energy splitting of these magnetic impurities is largely determined by the effective field from the host lattice, \textit{i.e.}, the exchange splitting, $\omega_{ex}$. These splittings determine whether resonance or relaxation dominates the magnon dissipation.

 The temperature dependence shown in Fig.~\ref{fig:fig4} suggests that around 8 K, the magnon damping is dominated by the relaxation mode associated with magnetic impurities that have a relatively large energy splitting, whereas below 2 K, there appear to be magnetic impurities with small energy splittings comparable to $\omega_0$; this leads to the dominance of the resonant mode. In Fig.~\ref{fig:fig4}, the solid line represents the calculated temperature profile based on the TLS model that incorporates both relaxation and resonant modes: $\Delta H (T) =\Delta H^{res.}_{TLS} (T,\omega_{ex,1}) + \Delta H^{relax.}_{TLS} (T,\omega_{ex,2})$, where $\omega_{ex,1}$ and $\omega_{ex,2}$ denote the exchange splittings of the TLS involved in the relaxation and resonant modes, respectively. Using the above two-TLS model, we extrapolate exchange splittings, $\omega_{ex,1}=26.8$ GHz (1.3 K) and $\omega_{ex,2}=296$ GHz (14.2 K), for the resonant and relaxation modes, respectively. The two-TLS model fits well across the various FMR frequency $\omega_0$, although shifts in $\omega_0$ slightly modify the energy splittings of the TLS, see Appendix~\ref{Appendix:TLS Model}. These extrapolated exchange splittings support the existence of two distinct types of TLS, each of which dominates the FMR broadening in different temperature regimes.  We also examine the impact of distributed TLS exchange splittings on the FMR linewidth, since aluminum doping can introduce random fluctuations in the local chemical environment of the TLS. We find that assuming a flat distribution of exchange splittings does not produce substantial changes in the temperature dependence of the FMR linewidth. Therefore, we retain our fixed-exchange-splitting model to fit the data.

\section{Discussion}
\label{MS:Discussion}

\begin{figure}[h!]
\centering
\includegraphics[width=\columnwidth]{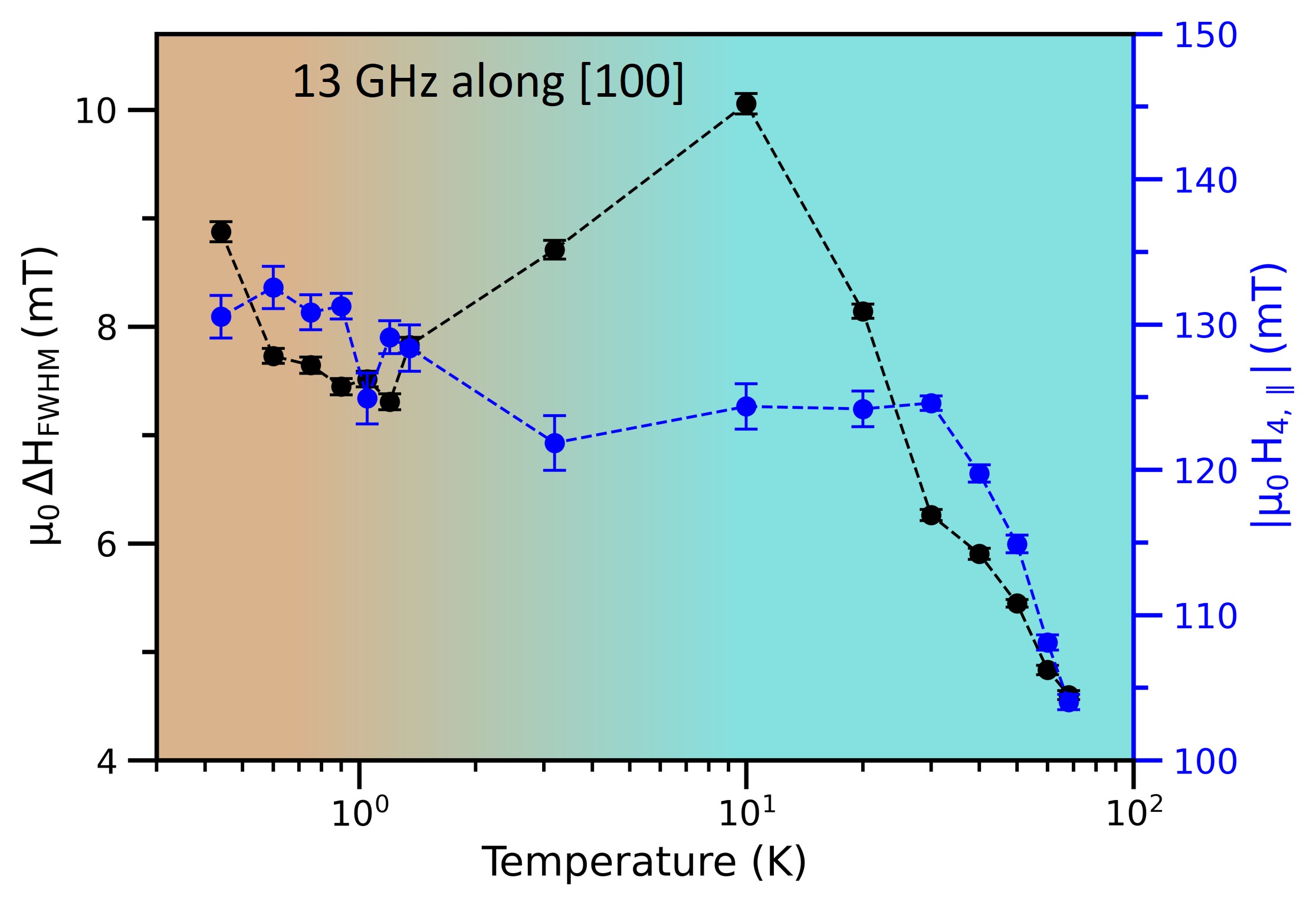}
\caption{Correlation between linewidth and anisotropy with the field along the [100] direction at 13 GHz. Black symbols (left axis) show the temperature dependence of the FMR linewidth, while blue symbols (right axis) show the corresponding evolution of the 4-fold in-plane the anisotropy field. The shaded background serves as a visual guide to temperature, highlighting the crossover from a low-temperature regime dominated by paramagnetic impurity spins to a high-temperature regime where exchange-coupled impurities become relevant, as discussed in Section~\ref{MS:Discussion}. Error bars represent the standard errors obtained from nonlinear least-squares fits to the frequency and linewidth data.} 
\label{fig:fig5}
\end{figure}

Our broadband FMR measurements reveal three main features that together shape the low-temperature spin dynamics of LAFO thin films.  First, the linewidth exhibits a nonmonotonic temperature dependence, peaking near 8 K and increasing subtly below 2 K.  Second, the magnetic anisotropy undergoes significant modification at low temperature, including the emergence of an unusually large sixth-order cubic term along the [110] direction.  Third, the temperature evolution of the anisotropy coefficients extracted along [100] shows trends that parallel the linewidth behavior (Fig.~\ref{fig:fig5}). These observations indicate that the same microscopic mechanisms govern both damping and anisotropy, with the temperature-dependent coupling between defect spins and the ferrimagnetic lattice playing a central role.

Within the TLS-based theoretical framework, magnon dissipation arises from impurity spins that possess an energy splitting $\omega_{\mathrm{imp}}$ and couple to the host lattice via anisotropic exchange. The experimental behavior can be understood by considering the coexistence of at least two classes of magnetic defects with distinct characteristic energy splittings. One class, predominantly active at higher temperatures, consists of \emph{exchange-coupled impurities}--such as Fe$^{3+}$ moments partially integrated into the ferrimagnetic backbone through Fe--O--Fe or Fe--O--Al pathways---have splittings much larger than the FMR frequency, due to the existence of the large molecular field associated with the host spins. These defects contribute primarily through a relaxation-type process giving rise to the broad linewidth maximum near $\sim$8 K. A second class of defects, dominant at lower temperatures, consists of more weakly coupled defect spins that behave as nearly free \emph{paramagnetic centers} with comparatively small energy splittings set by their local chemical environments rather than by temperature. At higher temperatures, these small-splitting TLS are thermally saturated and therefore do not contribute appreciably to magnon damping, as their ground and excited states are nearly equally populated. Upon cooling, however, these defects become unsaturated and resonant with the FMR frequency, allowing them to damp the precession efficiently through a resonant absorption process. This mechanism naturally accounts for the slight upturn in the linewidth observed below $\sim$2 K. The \emph{coexistence} of two types of magnetic impurities with distinct energy splittings thus provides a consistent explanation for the non-monotonic temperature dependence of the FMR linewidth, without requiring individual defects to modify their intrinsic coupling strengths or energy scales upon cooling.

An additional or complementary factor may involve subtle lattice modifications at cryogenic temperatures, arising either from a structural or isostructural transition or from strain induced by differential thermal expansion between the film and the substrate. By structural transition, we refer to a change in crystallographic symmetry, whereas an isostructural transition denotes a modification of internal lattice parameters—such as bond lengths, bond angles, or local distortions—without a change in the global space group. Similarly, thermally induced strain can continuously modify internal lattice parameters through elastic deformation. The change in the temperature dependence of the anisotropy fields at cryogenic temperatures suggests the presence of such lattice modifications, which  can significantly modify both the lattice constant and exchange interaction within the host lattice~\cite{Liu2025}. Such modifications alter the local chemical environment surrounding the magnetic impurities, thereby reducing the effective fields acting on certain impurities, transforming them from exchange-coupled to weakly coupled states. In this scenario, the two types of impurities inferred from the previous linewidth analysis would correspond to the same defect species residing in distinct local environments, whose relative populations and coupling strengths evolve as the lattice changes.

Two microscopic interpretations are therefore consistent with our data. In a \emph{coexistence model}, exchange-coupled and paramagnetic impurities form distinct populations that dominate different temperature ranges. In a \emph{transition model}, a single defect species changes its coupling strength to the host lattice, and hence its energy splitting, potentially due to lattice modifications. Both scenarios reproduce the temperature dependence of the linewidth, and they may in practice coexist.

The anisotropy coefficients and the FMR linewidth exhibit a correlated temperature dependence, as summarized in Fig.~\ref{fig:fig5}. Within the framework developed above, such correlated behavior can arise in two distinct but physically consistent ways. In the coexistence picture, although randomly oriented impurity spins would not by themselves produce a net magnetic anisotropy, even weak coupling to the crystalline host through spin–orbit interactions or exchange pathways can bias the spin quantization axes of the impurity spins according to the lattice symmetry. As a result, defect spins can inherit the host anisotropy and contribute to an effective, symmetry-allowed anisotropy rather than averaging to zero. Alternatively, in the transition picture, the lattice evolution that drives the transformation of defects from exchange-coupled to more weakly coupled states also influences the magnetic anisotropy via magnetoelastic coupling, leading to correlations in the temperature dependence of the magnetic anisotropy and linewidth.

The fourfold in-plane anisotropy $H_{4,\parallel}$ increases steadily as the temperature is reduced from 68 K to approximately 30 K, over the same temperature range in which the linewidth also grows. This concurrent evolution can arise either from the increasing influence of exchange-coupled defects in the coexistence picture, or from gradual lattice-driven modifications of defect environments in the transition picture. Below $\sim$30 K, $H_{4,\parallel}$ enters a plateau regime, while the linewidth continues to increase and reaches a broad maximum near $\sim$8 K. In the coexistence picture, this reflects a regime in which exchange-coupled defects dominate the anisotropy but remain dynamically active as relaxation channels, whereas in the transition picture the anisotropy has largely stabilized while defect-mediated dissipation continues to evolve. At temperatures below $\sim$2 K, both $H_{4,\parallel}$ and the linewidth show a subtle enhancement with further cooling, consistent with the growing influence of weakly coupled or paramagnetic centers in the coexistence picture, or with continued lattice-driven evolution of defect environments in the lattice-modification picture.

The uniaxial anisotropy $H_{2,\perp}$ shows similar trends over the same temperature range as can be seen in its temperature evolution in Fig.~\ref{fig:fig3}(b). Although it has not been explicitly plotted alongside the linewidth, its crossover temperature aligns with that inferred from the $H_{4,||}$ behavior as can be seen in Fig.~\ref{fig:fig3}(a)-(b), suggesting that the same TLS and exchange-coupled impurity mechanisms influencing the linewidth also subtly affect the uniaxial component. This provides additional, independent support for the picture of two types of impurities in LAFO.

The unusually large sixth-order cubic anisotropy observed along [110] can be understood as a consequence of the same microscopic processes discussed above. Since the [110] direction is more sensitive to higher-order cubic harmonics, any local symmetry breaking or exchange imbalance amplifies these terms. The presence of such a large higher-order anisotropy produces a more complex FMR energy landscape along [110], increasing sensitivity to spatial inhomogeneity and allowing multiple quasi-degenerate resonance conditions. As a result, the FMR spectra measured along [110] exhibit an increased linewidth compared to the response observed along the [100] direction.

Future temperature-dependent Raman or high-resolution X-ray diffraction studies, together with local magnetic probes such as NMR, XMCD or M\"{o}ssbauer spectroscopy, will be essential to determine whether the magnetic crossover is accompanied by a structural or isotructural transition. Regardless of the microscopic origin, these results establish that at cryogenic temperatures, spin coherence in LAFO is limited by defect-induced TLS that couple to magnons through both relaxation and resonant processes, with possible reinforcement from subtle structural rearrangements in the spinel lattice.
  
\section{Conclusions}
\label{MS:Conclusion}

Our broadband FMR study of LAFO thin films uncovers a crossover in spin dynamics from cubic anisotropy--dominated behavior to TLS-limited coherence at cryogenic temperatures. The [100] resonance is captured by conventional four-fold and uniaxial anisotropy fields, while the [110] axis requires a significant sixth-order cubic anisotropy at cryogenic temperatures. The temperature evolution of the linewidth, with a peak near 8 K and a slight upturn below 2 K, points to dissipation mediated by TLS, consistent with contributions from both exchange-coupled spins and isolated paramagnetic centers. These results demonstrate that intrinsic coherence in LAFO is fundamentally limited by antisite defects that couple incoherently to magnons at ultralow temperatures. More broadly, they establish TLS as an important relaxation channel in ferrimagnetic insulators, with implications for the design of low-loss materials for magnonics, quantum transduction, and hybrid spin--photon devices.

\section{Acknowledgments}
\label{MS:Ackn}

This work was supported as part of the Center for Energy Efficient Magnonics an Energy Frontier Research Center funded by the U.S. Department of Energy, Office of Science, Basic Energy Sciences at SLAC National Laboratory under contract \# DE-AC02-76SF00515. Cryogenic measurements were enabled by the cryostat facility of the Center for Molecular Quantum Transduction, an Energy Frontier Research Center funded by the U.S. Department of Energy, Office of Science, Office of Basic Energy Sciences, under Award No. DE-SC0021314.

\section{Author Contributions}
\label{MS:Auth}

S.P, G.D.F., C.C.Y, and Y.S. conceived the project. S.P. fabricated the Ti/Cu/Pt broadband waveguides, carried out the low-temperature broadband FMR measurements, and analyzed the data. G.Q. and H.M.C. performed the theoretical calculations. K.M. and L.T. performed the sample growth and additional characterization (SQUID and room-temperature broadband FMR). Q.X. did the magnetic field calibration and assisted in low-temperature FMR measurements. S.P. prepared the manuscript with inputs from all the co-authors. Y.S., C.C.Y. and G.D.F. supervised the project.

\appendix

\section{MATERIAL CHARACTERIZATION}
\label{Appendix:Mat Char}

\begin{figure}[h!]
\centering
\includegraphics[width=\columnwidth]{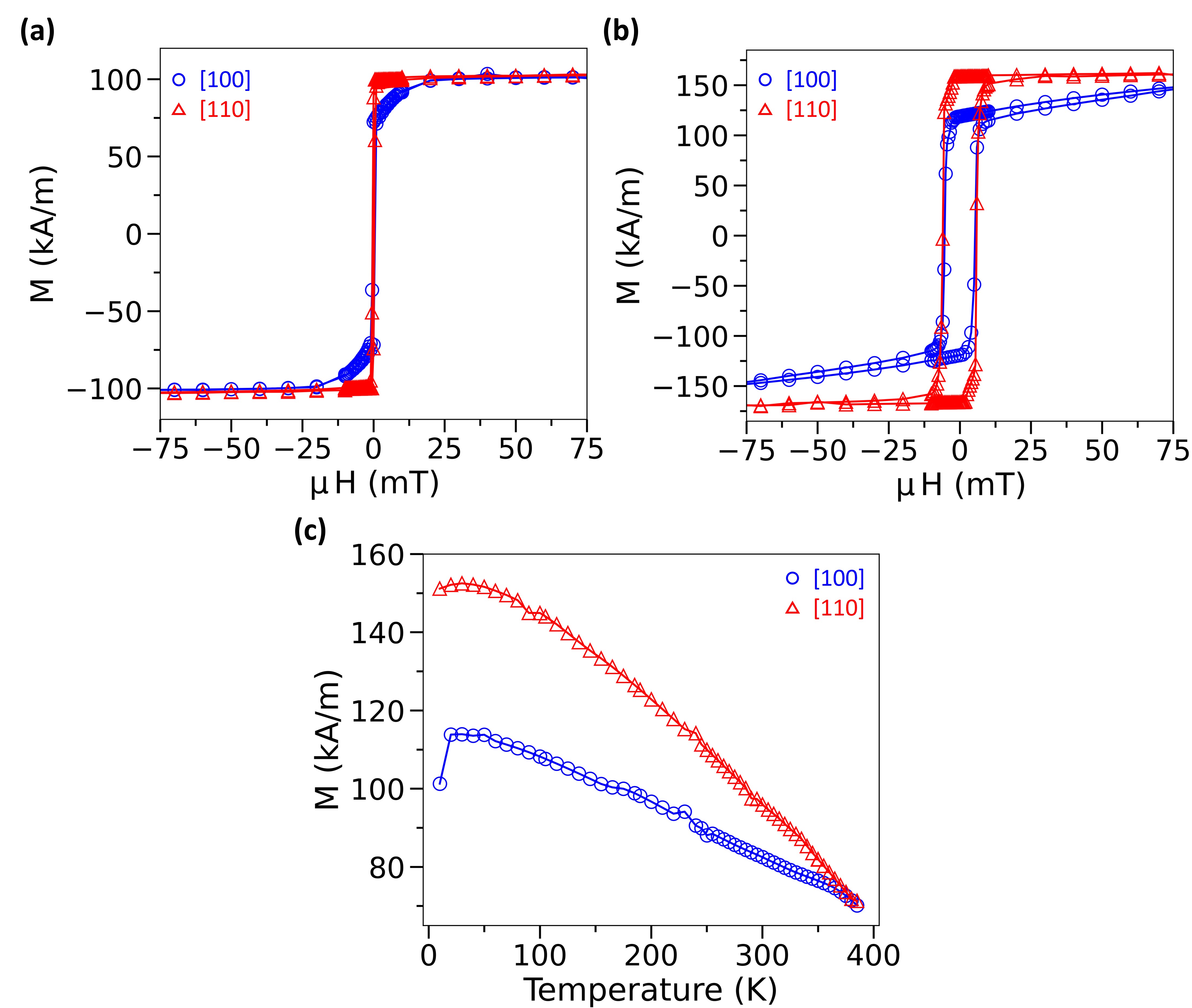}
\caption{(a)-(b) SQUID Hysteresis loops along [100] and [110] at 300 K and 10 K, respectively. (c) Temperature dependence of extracted saturation magnetization along [100] and [110] directions.}
\label{fig:fig6}
\end{figure}

To characterize the static magnetic properties, we use the Quantum Design Evercool SQUID magnetometer. We study static magnetization as a function of the magnetic field applied along both the [100] and [110] in-plane axes at 300 K and 10 K, and the results are shown in Fig.~\ref{fig:fig6}(a) and (b), respectively. At both 300 K and 10 K, the [110] hysteresis loop is square relative to the [100] hysteresis loop, which indicates that the [110] direction is the easy axis. In Fig.~\ref{fig:fig6}(c), the magnetization along [110] remains higher than [100] as the temperature is decreased from 300 K to 10 K showing that [110] is the easy axis. The magnetization at 10 K along the [110] axis is 150 kA/m which is similar to the saturation magnetization observed in Fig.~\ref{fig:fig6}(b). Collectively, these results demonstrate that [110] is the easy axis.

\section{BROADBAND WAVEGUIDE FABRICATION}
\label{Appendix:Waveguide Fab}

The broadband waveguides were fabricated using standard photolithography on sapphire substrates. Initially, the wafers were cleaned via sonication in acetone followed by isopropyl alcohol (IPA) and then coated with a bilayer of LOR5A and S1813 photoresist. The coated wafers were exposed using a 5× g-line wafer stepper and developed in AZ726MIF. After development, the samples were descummed and a 225 nm Ti/Cu/Pt tri-layer was deposited using an AJA sputter deposition system. Finally, the unwanted metal was removed via lift-off in MicroChem 1165.

\section{LOW-TEMPERATURE FMR MEASUREMENT SCHEMATIC}
\label{Appendix:Schematic}

\begin{figure}[h!]
\centering
\includegraphics[width=50 mm]{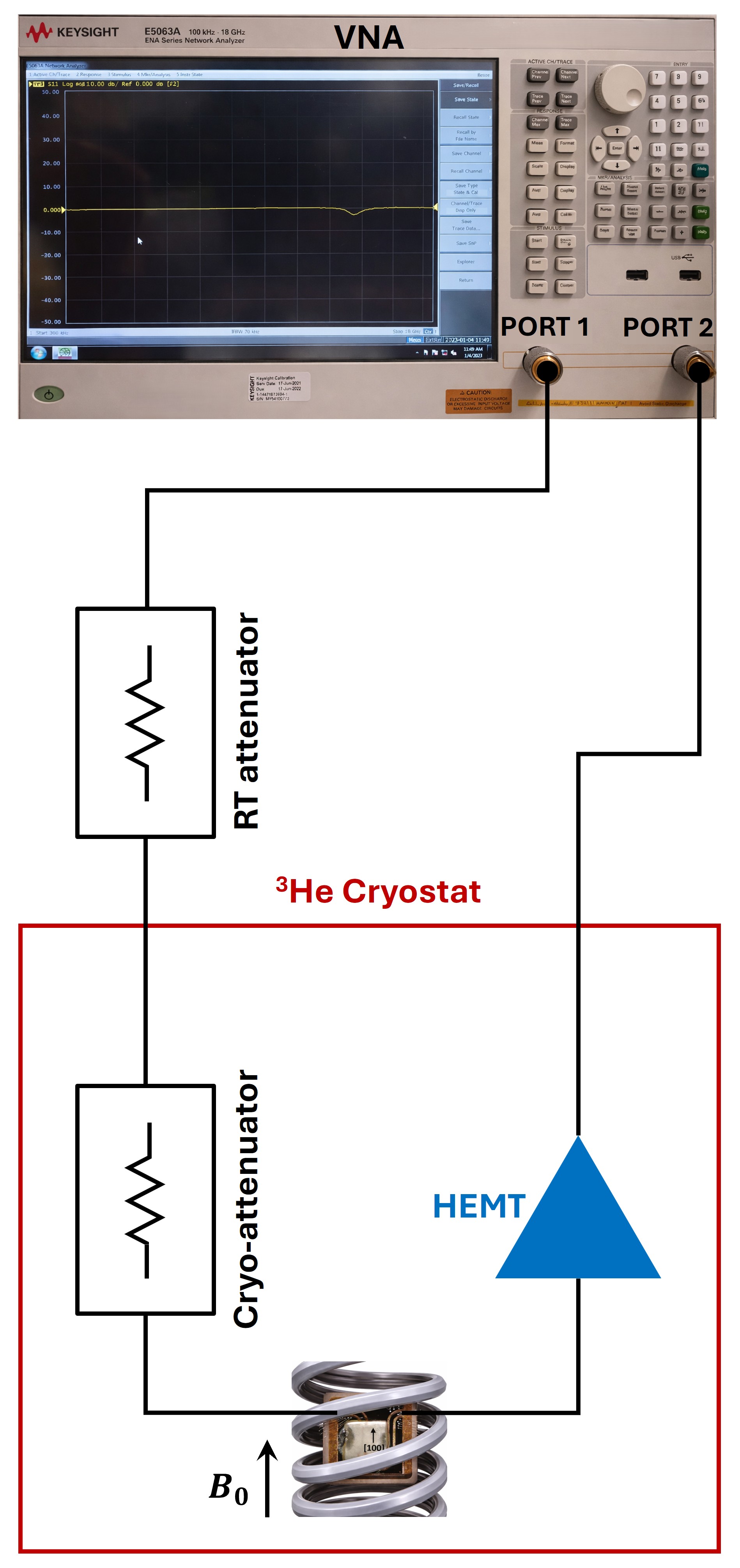}
\caption{Schematic of the low-temperature ferromagnetic resonance (LT-FMR) measurement setup.} 
\label{fig:fig7}
\end{figure}

Figure~\ref{fig:fig7} shows a schematic of the microwave measurement setup used for low-temperature ferromagnetic resonance (LT-FMR) experiments. A vector network analyzer (VNA) provides the microwave excitation and measures the complex scattering parameters of the device under test in a two-port configuration.

The microwave signal from Port 1 of the VNA is routed through room-temperature (RT) attenuators and through additional cryogenic attenuators mounted at different cold stages inside the cryostat, which thermalizes the microwave line and reduces thermal noise incident on the sample. The attenuated microwave excitation is then delivered to the sample, which is placed inside a superconducting electromagnet used to generate a static magnetic field $B_0$ along the indicated direction.

The transmitted microwave signal from the sample is routed back through the output line and amplified using a cryogenic high-electron-mobility transistor (HEMT) amplifier mounted at low temperature (3 K) to achieve low-noise signal amplification and then is fed to Port 2 of the VNA for detection.

All microwave components are impedance-matched to 50 $\Omega$ to minimize reflections and standing waves. The use of distributed attenuation and low-noise amplification enables stable, broadband measurements of the FMR response as a function of frequency, magnetic field, and temperature.

\section{ABSORBED POWER AT RESONANCE}
\label{Appendix:P_res}

A notable observation from the low-temperature FMR spectrum as shown in Fig.~\ref{fig:fig1}(c)-(d) is the enhanced noise level in the spectra recorded with the field along the [110] direction compared to those along [100]. To verify that this difference is intrinsic to the magnetic response rather than an experimental artifact arising from imperfect sample contact or mounting, we quantified the resonance strength using the magnitude of the absorptive Lorentzian amplitude $|L|$ extracted from fits to the transmission spectra since that is the parameter proportional to the microwave power absorbed by the sample. The dispersive Lorentzian component $D$, which arises from phase mixing and impedance mismatch in the transmission geometry, does not contribute to the absorbed power and therefore is not relevant for quantifying the resonance strength. The $|L|$ parameters extracted from the red fits in Fig.~\ref{fig:fig1} (c)-(d) along [100] and [110] directions are 0.0096 (5) dBm and 0.0106 (2) dBm, respectively. The comparable values of $|L|$ for the two crystallographic orientations indicate that the observed noise difference cannot be attributed to variations in sample contact or coupling conditions.

\section{ROOM-TEMPERATURE BROADBAND FMR MEASUREMENTS}
\label{Appendix:RT FMR}

To characterize the dynamic properties, we conduct room-temperature broadband FMR measurements using a coplanar waveguide geometry. The measurements were performed along the [100] and [110] in-plane axes for a range of frequencies from 5 GHz to 29 GHz. Figure~\ref{fig:fig7} shows the frequency dependencies of the FMR field and the linewidth. In Fig.~\ref{fig:fig7}(a), the solid lines are fits to Eq. \ref{eq:SB_Fit} with $H_{6,||}=0$. We see the expected lower FMR field at each frequency along the [110] axis as compared to the [100] axis. In Fig.~\ref{fig:fig7}(b), the solid lines are fits to Eq.~\ref{eq:linewidth_model} with the third term negligible within the experimental uncertainty. From these fits, we extract a damping of $\alpha = 1.1(1)\times10^{-3}$ along the [100] axis and a damping of $\alpha = 0.6(1)\times10^{-3}$ along the [110] axis.

\begin{figure}[h!]
\centering
\includegraphics[width=\columnwidth]{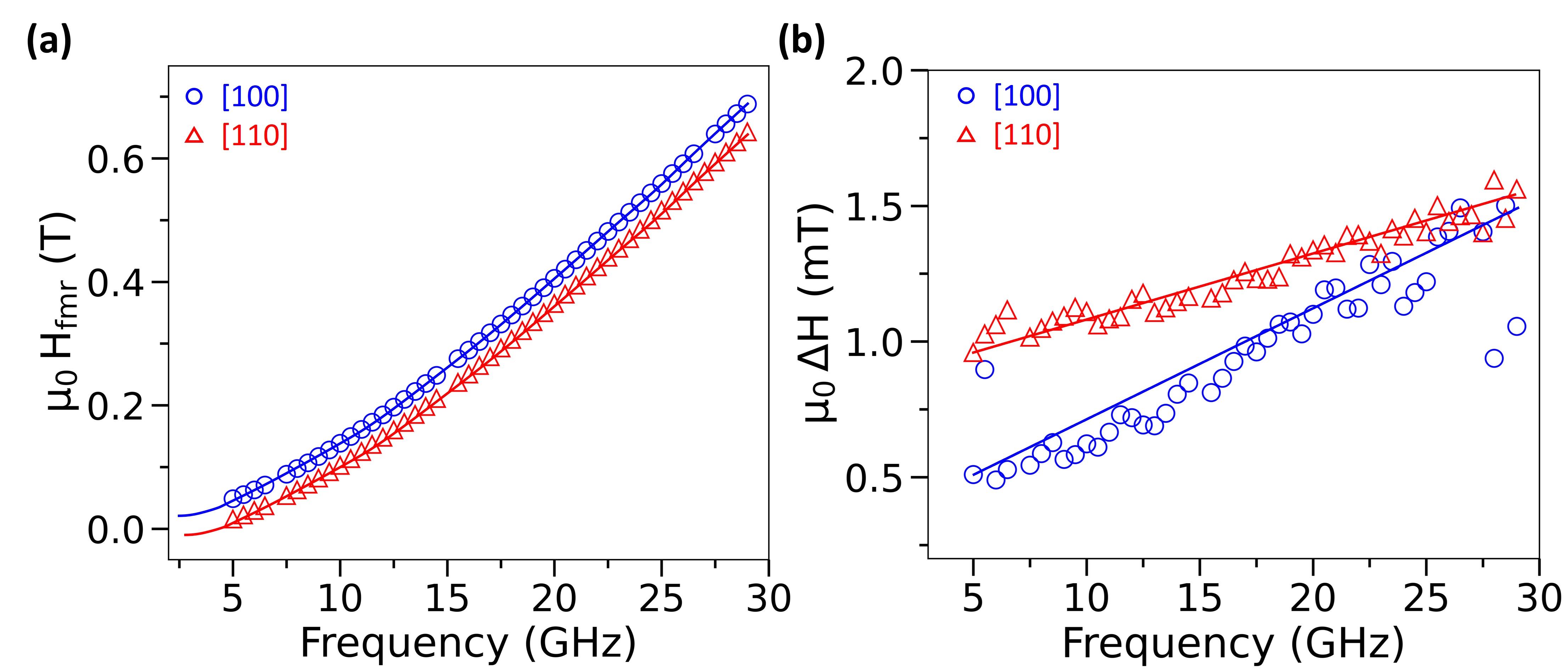}
\caption{(a)-(b) Frequency variation of room-temperature FMR resonance field and linewidth, respectively. Blue data corresponds to the magnetic field applied along the [100] direction and red data corresponds to the field applied along the [110] direction.} 
\label{fig:fig8}
\end{figure}

\section{TEMPERATURE DEPENDENT FMR RESPONSE ALONG [110]}
\label{Appendix:T-dep FMR [110]}

\begin{figure}
\centering
\includegraphics[width=\columnwidth]{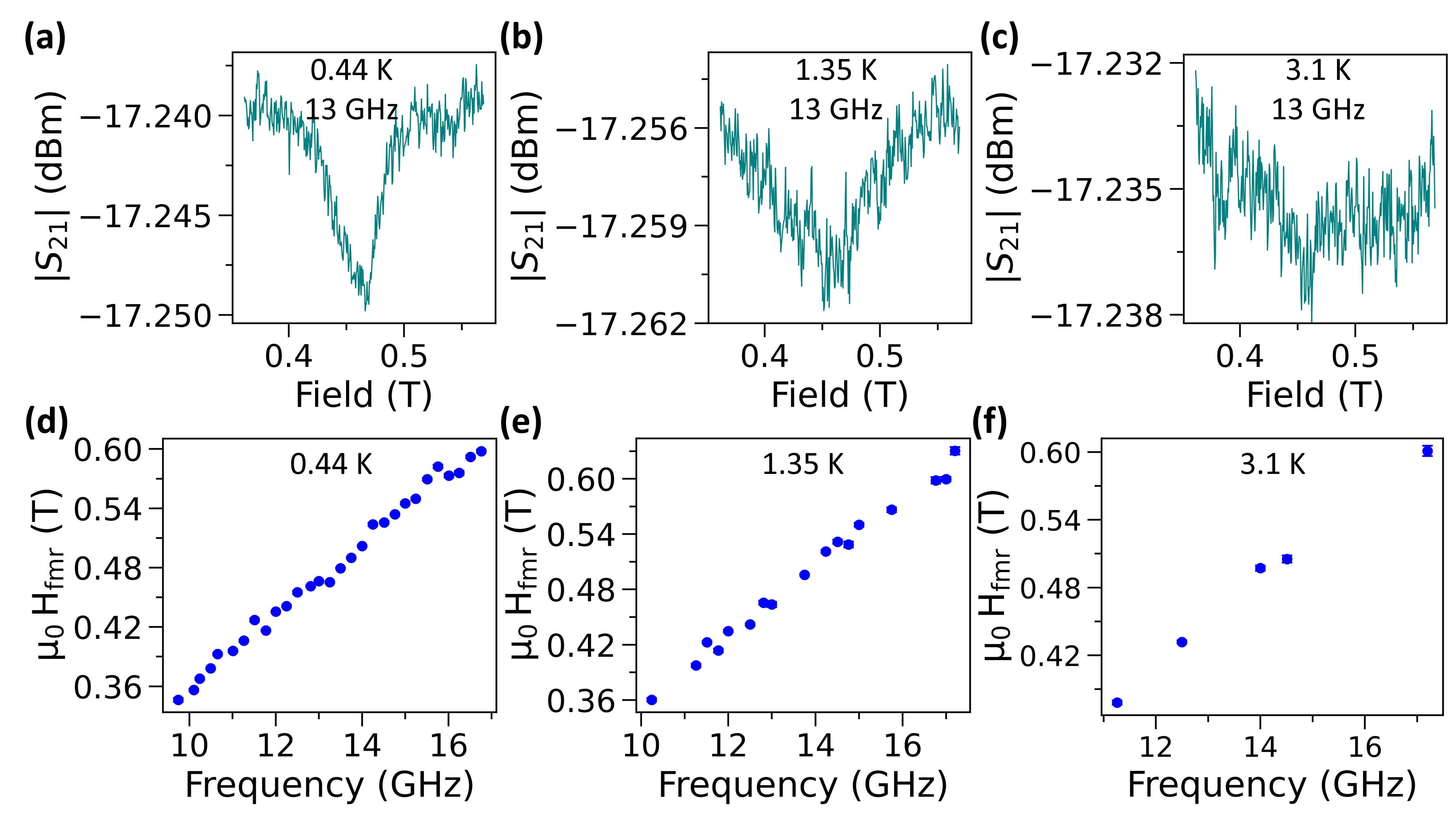}
\caption{(a)-(c) Frequency dependence of resonance field along [110] at 0.44 K, 1.35 K, and 3.2 K, respectively. (d)-(f) FMR spectra at 13 GHz along [110] at 0.44 K, 1.35 K, and 3.1 K, respectively.}
\label{fig:fig9}
\end{figure}

As shown in Fig.~\ref{fig:fig1}(c)-(d) and discussed in the main text, the FMR transmission along [110] is noticeably noisier than along [100] at 0.44 K. To track the temperature dependence of the [110] response, we measured the spectra at different temperatures. With increasing temperature, the spectra broaden and the signal strength decreases significantly. Figure~\ref{fig:fig9}(a)-(c) shows the transmission spectra along [110] at 13 GHz for 0.44 K, 1.35 K, and 3.1 K. The data clearly show progressive broadening and a reduced resonance depth, making the signal unsuitable for fitting at 3.1 K. A similar trend was observed at other measurement frequencies as well: with increasing temperature, spectra that could be reliably fitted became progressively fewer. This is reflected in the frequency–field dispersion plot (Fig.~\ref{fig:fig9}(d)-(f)), where the reduction in the number of data points directly corresponds to spectra with signal-to-noise ratios too low for meaningful fitting.

\section{TLS CONTRIBUTION TO LINEWIDTH}
\label{Appendix:TLS Model}

The isolated Fe cations serve as two-level magnetic impurities that induce magnon dissipation through two distinct channels: the resonant and relaxation processes. The interaction between a magnetic impurity and the Fe host lattice can be written as
\begin{align}
\mathcal{H}_{\text{int}} = - c_{imp} \lambda M_{\alpha} J_{\alpha \beta} m_{\beta},
\label{eq:Appen-F-1}
\end{align}
where $\bm{M}$ and $\bm{m}$ denote the magnetization of the host lattice and the magnetic impurity, respectively. The parameter $c_{imp}$ represents the impurity concentration, and $\lambda = V_c/(g\mu_B)^2$ is the coupling strength, with $V_c$ being the cluster volume surrounding a single magnetic impurity. We assume that the exchange coupling is uniaxial, described by the matrix $\bm{J} = \mathrm{diag}(J_{\perp}, J_{\perp}, J_{\parallel})$, where the $\hat{z}$ direction defines the quantization axis.

The  Landau–Lifshitz–Gilbert (LLG) equations for both the host lattice magnetization and the magnetization of an impurity are
\begin{align}
\label{eq:Appen-F-2}
\frac{ d M_{\gamma} }{ d t} &=- \gamma \varepsilon_{ \alpha \beta \gamma} M_{\alpha} \left( B_{0,\beta} + c_{imp} \lambda   J_{\beta \tau} m_{\tau} \right), \\
\frac{ d m_{\gamma} }{ d t} &= - \gamma \varepsilon_{ \alpha \beta \gamma} m_{\alpha} \left( B_{0,\beta} + \lambda   J_{ \tau \beta} M_{\tau} \right) ,
\label{eq:Appen-F-3}
\end{align}
which are obtained from the Hamiltonians for host lattice and magnetic impurity separately,
\begin{align}
\label{eq:Appen-F-14}
H_{\text{host}} &=   - M_{\alpha}  \bm{B}_{\text{eff},\alpha} - c_{imp}  \lambda M_{\alpha} J_{\alpha \beta} m_{\beta}, \\
H_{d} &=   - m_{\alpha}  \bm{B}_{\text{eff},\alpha} -  \lambda M_{\alpha} J_{\alpha \beta} m_{\beta}.
\label{eq:Appen-F-5}
\end{align}
Here, $\bm{B}_{0} = (0, 0, B_{0})$ denotes the external magnetic field. Note that the magnetic anisotropy within the host lattice is neglected.

The magnetic impurities can be treated as an ensemble of TLS coupled to the host lattice. Their energy splitting is determined by the Zeeman energy given in Eq.~\eqref{eq:Appen-F-5},
\begin{align}
\varepsilon_{\pm} = \pm  \frac{\hbar \omega_{imp}}{2} = \pm \frac{\hbar}{2} \gamma_g \left| \sum_\beta B_{0,\beta} +B_{d,\beta} \right|, 
\label{eq:Appen-F-6}
\end{align}
where $\gamma_g$ is the gyromagnetic ratio, and $\bm{B}_d = \lambda \bm{J}^{\mathrm{T}}\bm{M}$ represents the molecular field of the host lattice acting on the impurity spins. If the external field and molecular field are collinear (\textit{i.e.}, the interaction $\bm{J}$ is diagonal), the energy splittings of impurities can be separated into $\omega_{ \rm{imp}} \approx \omega_{0} + \omega_{ex}$, where $\omega_0$ is the FMR frequency and $\omega_{ex}=\gamma |\bm{B}_d|$ is the exchange splitting induced by the exchange interaction with the host lattice.

At finite temperature, the magnetization of impurity spins in thermal equilibrium reads
\begin{eqnarray}
m_{\text{eq}} (T) = m_0 \tanh \frac{ \hbar \omega_{imp} }{2 k_B T},
\label{eq:Appen-F-7}
\end{eqnarray}
where $m_0$ is the saturation magnetization of impurity spins.

\subsection{Relaxation mode: longitudinal relaxation of the impurity spins}
In the relaxation mode, the impurity spins relax toward thermal equilibrium with a characteristic longitudinal time $\tau_{imp}$. In this scenario, the precessional motion in the LLG, which is purely transverse [Eq.~\eqref{eq:Appen-F-5}],  can be neglected, reducing it to a Bloch-type relaxation equation along the longitudinal direction,
\begin{eqnarray}
\frac{d \bm{m} }{ d t}  = - \frac{1}{\tau_{imp}} \left(  \bm{m} \cdot \hat{e} - m_{\text{eq}} (T)   \right)\hat{e} ,
\label{eq:Appen-F-9}
\end{eqnarray}
where $\hat{e} = \left( \bm{B}_0 + \bm{B}_d \right) / |\bm{B}_0 + \bm{B}_d|$ denotes the direction of the total field of acting on the impurity spins. When uniaxial exchange coupling dominates, we have $\hat{e} = \hat{z}$, as the transverse component is relatively small.

The transverse perturbation of the host-lattice magnetization, $\bm{M}_{\perp}(t)$, induces a variation in the molecular field acting on the impurity spins, $\delta \bm{B}_d = \lambda \bm{J}^{\mathrm{T}}\bm{M}_{\perp}$. Consequently, this perturbation drives the TLS toward a new thermal-equilibrium configuration,
\begin{align}
 \delta m_{\text{eq}} (T) =  \frac{ \partial m_{\text{eq}} (T)  }{ \partial B_d} \delta B_{||} &=  \chi_{0,||} \delta B_{||}, \notag \\
 \chi_{0,||}&= \frac{m_0 \hbar \gamma}{2 k_B T} \text{sech}^{2} \frac{ \hbar \omega_{imp} }{2 k_B T}.
\label{eq:Appen-F-10}
\end{align}

The longitudinal and transverse components of perturbed impurity spins are
\begin{eqnarray}
\delta m_{||} =  \frac{\chi_{0,||} }{ 1 - i \omega \tau_{imp}}  \delta B_{||},  \quad \delta \bm{m}_{\perp} = \frac{  m_{\text{eq}} (T) }{| \bm{B}_d|} \delta \bm{B}_{\perp},
\label{eq:Appen-F-11}
\end{eqnarray}
where the transverse component of the perturbed effective field, $\delta \bm{B}_{d,\perp} = P_{\perp}\delta \bm{B}_d$, only induces an instantaneous tilt of the impurity magnetization. 

By combining the longitudinal and transverse responses, the dynamic susceptibility of the magnetic impurities can be written as
\begin{align}
\delta \bm{m} &= \bm{\chi}_d\delta \bm{B}_d , \notag \\
\bm{\chi}_d & = \frac{\chi_{0,||} }{ 1 - i \omega \tau_{imp}} \hat{e} \hat{e}^T  + \frac{ m_{\text{eq}} (T) }{| \bm{B}_d|} \left(  \bm{I} - \hat{e} \hat{e}^T \right) .
\label{eq:Appen-F-12}
\end{align}

By replacing the impurity spin in Eq.~\eqref{eq:Appen-F-2} with its dynamic susceptibility, $\bm{\chi}_d$, the interaction term can be projected back onto the magnetization of the host lattice.
\begin{equation}
\frac{ d \bm{M} }{ d t} = - \gamma \bm{M} \times \left( \bm{B}_{0}  + c_{imp} \lambda^2  \left(\bm{J} \bm{\chi}_d \bm{J}^T \right)  \bm{M}  \right).
\label{eq:Appen-F-13}
\end{equation}

By linearizing the LLG equation, we obtain the resonance frequency,
\begin{equation}
\begin{aligned}
\omega
&= \gamma B_{0}
+ c_{\mathrm{imp}}\,\lambda^{2}\gamma M_{0}\Bigg(
\frac{\chi_{0,\parallel}}{1-i\omega\tau_{\mathrm{imp}}}\,
\frac{J_{\parallel}^{4}-J_{\perp}^{4}}{J_{\parallel}^{2}}
\\
&\qquad\qquad\qquad\qquad
-\frac{m_{\mathrm{eq}}(T)}{\lambda J_{\parallel} M_{0}}\,
\frac{J_{\perp}^{2}\!\left(J_{\parallel}^{2}-J_{\perp}^{2}\right)}{J_{\parallel}^{2}}
\Bigg),
\end{aligned}
\label{eq:Appen-F-14}
\end{equation}
where the imaginary part gives the FMR linewidth,

\begin{align}
\Delta \omega & =  \frac{c_{imp} m_0}{M_0}   \frac{ J^4_{||}-J^4_{\perp}}{J^4_{||}} \frac{ \hbar \omega^2_{ex} }{2 k_B T } \frac{  \omega_0 \tau_{imp} }{ 1 + \omega_0^2 \tau^2_{imp}}   \text{sech}^{2} \left( \frac{ \hbar \omega_{imp} }{2 k_B T}\right) .
\label{eq:Appen-F-15}
\end{align}

The linewidth of the field scan is obtained by a simple relation: $\mu_0 \Delta H = \partial   \omega /\partial B_0 \Delta \omega$.

\subsection{Resonant mode: impurity spins precess coherently with host lattice}
 
For the resonant mode, we consider the transverse relaxation of the impurity spin towards the instantaneous direction of \textcolor{blue}{the} impurity spins, $(m_{\text{eq}} (T) / |\bm{B}_0 + \bm{B}_d| ) \delta \bm{B}_{d,\perp}$.  
The transverse component of the LLG equation for the impurity spins can then be written as
\begin{align}
\frac{ d \bm{m}_{\perp} }{ d t}=&- \gamma \bm{m} \times \left( \bm{B}_0 + \bm{B}_d \right)\notag \\ &- \frac{1}{\tau_{imp}} \left(  \bm{m}_{\perp}  
- \frac{m_{\text{eq}} (T) }{|\bm{B}_0 + \bm{B}_d|} \delta \bm{B}_{d,\perp}  \right) .
\label{eq:Appen-F-16}
\end{align}

Defining the circular components of the transverse magnetization and the molecular field as $m_{\pm} = m_x \pm i m_y$ and $\delta B_{d,\pm} = \delta B_{d,x} \pm i \delta B_{d,y}$, respectively, the solution is
\begin{align}
m_{\pm}  
&= \chi_{\pm} M_{\pm} ,\notag \\
\chi_{\pm} & = \gamma  \lambda J_{\perp} \frac{ \pm 1 +  \frac{ i }{  \omega_{imp}  \tau_{imp}  }   }{   \left( \omega \pm \omega_{imp}  \right)  + \frac{i}{    \tau_{imp}} }  m_{\text{eq}} (T) .
\label{eq:Appen-F-17}
\end{align}
Note that the longitudinal component of the instantaneous impurity \textbf{}magnetization is, $m_{eq} (T) \hat{e}$.
Replacing the impurity spin with dynamic susceptibility $\chi_{\pm}$ and linearizing the LLG equation of the host lattice, we have the resonance frequency,
\begin{equation}
\begin{aligned}
\omega &= \gamma B_{0} + \gamma c_{imp} \lambda J_{||} m_{\text{eq}} (T)
\\
&\qquad
-\gamma^2 c_{imp} \lambda^2 J^2_{\perp} M_0 \frac{ \pm 1 + \frac{i}{\omega_{imp} \tau_{imp}}}{\left( \omega \pm \omega_{imp} \right) + \frac{i}{\tau_{imp}}} m_{\text{eq}} (T),
\end{aligned}
\label{q:Appen-F-18}
\end{equation}
where the linewidth is given by,
\begin{align}
\Delta \omega  
&=  \frac{ c_{imp} m_0}{M_0} \frac{ J^2_{\perp} }{  J^2_{||} } \frac{ \omega^2_{ex}}{\omega_{imp}} \frac{ \omega_{0} \tau_{imp} }{ 1 +(\omega_0 -\omega_{imp})^2 \tau^2_{imp}}  \tanh \frac{ \hbar \omega_{imp} }{2 k_B T} .
\label{q:Appen-F-19}
\end{align}

Note that in the above derivation, we assume a constant impurity relaxation time.

\subsection{Fitting of the FMR linewidth data}

\begin{figure}[h!]
\centering
\includegraphics[width=\columnwidth]{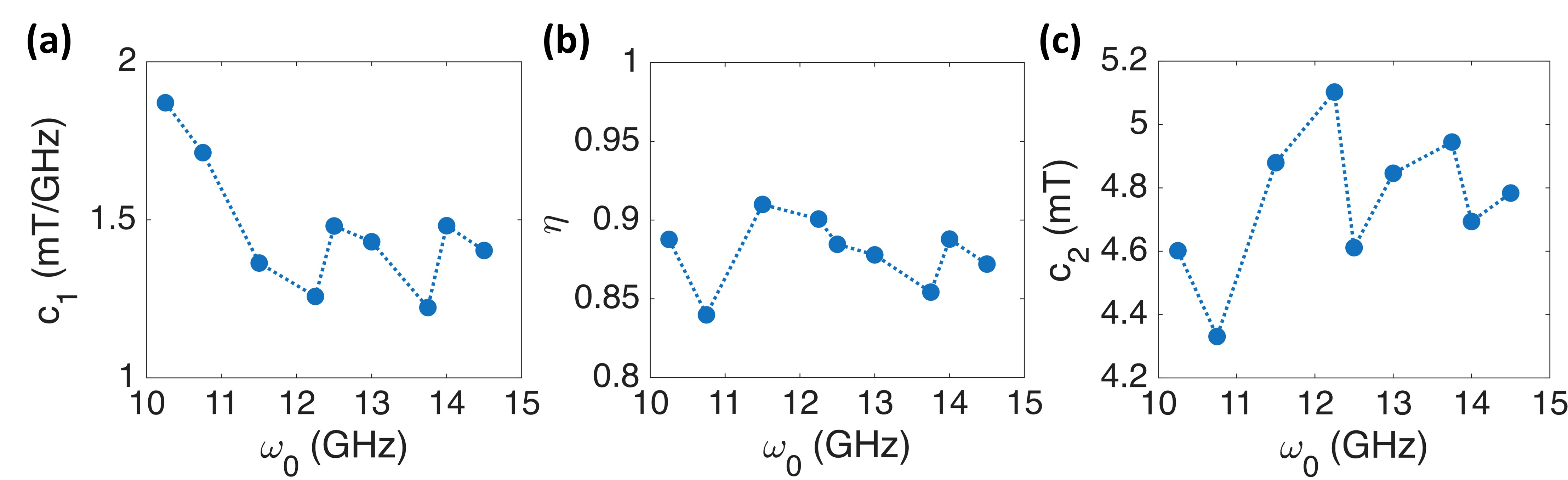}
\caption{Fitting parameters: (a) the prefactor $c_1$, (b) the exchange anisotropy $\eta$, and (c) the linewidth offset $c_2$, obtained with fixed exchange splittings $(\omega_{ex,1}, \omega_{ex,2}) = (26.8, 296)$\,\text{GHz}, at various FMR frequencies $\omega_0$. The relaxation time at $0$ K are set to be $\tau_{0,1}=100$ ps and $\tau_{0,2}=10$ ps for the resonant and relaxation, respectively.}
\label{fig:fig10}
\end{figure}

The FMR linewidth data sets at various resonance frequencies $\omega_0$ are fitted using a two-TLS model that incorporates both resonant and relaxation modes,
\begin{align}
\label{q:Appen-F-20}
    \Delta H_{total} (T) &= 
    c_1 \!\left[
        \eta^2\, F_{res}(T)
        + (1 - \eta^4)\, F_{rel}(T)
    \right]
    + c_2 ,
\end{align}
    where
    \begin{align}
F_{res} (T ) &=  \frac{ \omega_{0} \tau_{\rm{imp}} (T ) }{ 1 + \omega_{\rm{ex},1}^2  \tau^2_{\rm{imp}} (  T )  }  \frac{\omega^2_{{ex},1} }{ \omega_0 + \hbar \omega_{\rm{ex},1}} \tanh \left( \frac{ \hbar  \omega_0 + \hbar \omega_{\rm{ex},1}  }{2 k_B T} \right), \\
F_{rel} (T) &=  \frac{  \omega_0  \tau_{\rm{imp}} (  T )   }{ 1 + \omega_0^2  \tau^2_{\rm{imp}} (T)  }  \frac{ \hbar \omega^2_{\rm{ex},2} }{2 k_B T }  \cosh^{-2} \left( \frac{ \hbar  \omega_0 + \hbar \omega_{\rm{ex},2}  }{2 k_B T}\right) .
\label{q:Appen-F-21-22}
\end{align}
The fitting parameters are $c_1= \frac{ c_{imp} m_0}{\mu_0\gamma M_0} $, $\eta=\frac{J_{\perp} }{J_{||} }$, and $c_2$. $c_2$ represents the linewidth offset due to additional broadening mechanisms beyond the TLS contributions. $F_{res}$ and $F_{rel}$ are functions associated with the resonant and relaxation modes, respectively. Here, we incorporate the temperature dependent relaxation time, $\tau_{imp} (T)=\tau_0 \tanh \left(\hbar ( \omega_{0}+\omega_{ex})/2 k_B T \right)$.  For the zero-temperature relaxation times $\tau_0$, we set $\tau_{0,1} = 100$ ps and $\tau_{0,2} = 10$ ps for the resonant and relaxation modes, respectively.

\begin{figure}[h!]
\centering
\includegraphics[width=\columnwidth]{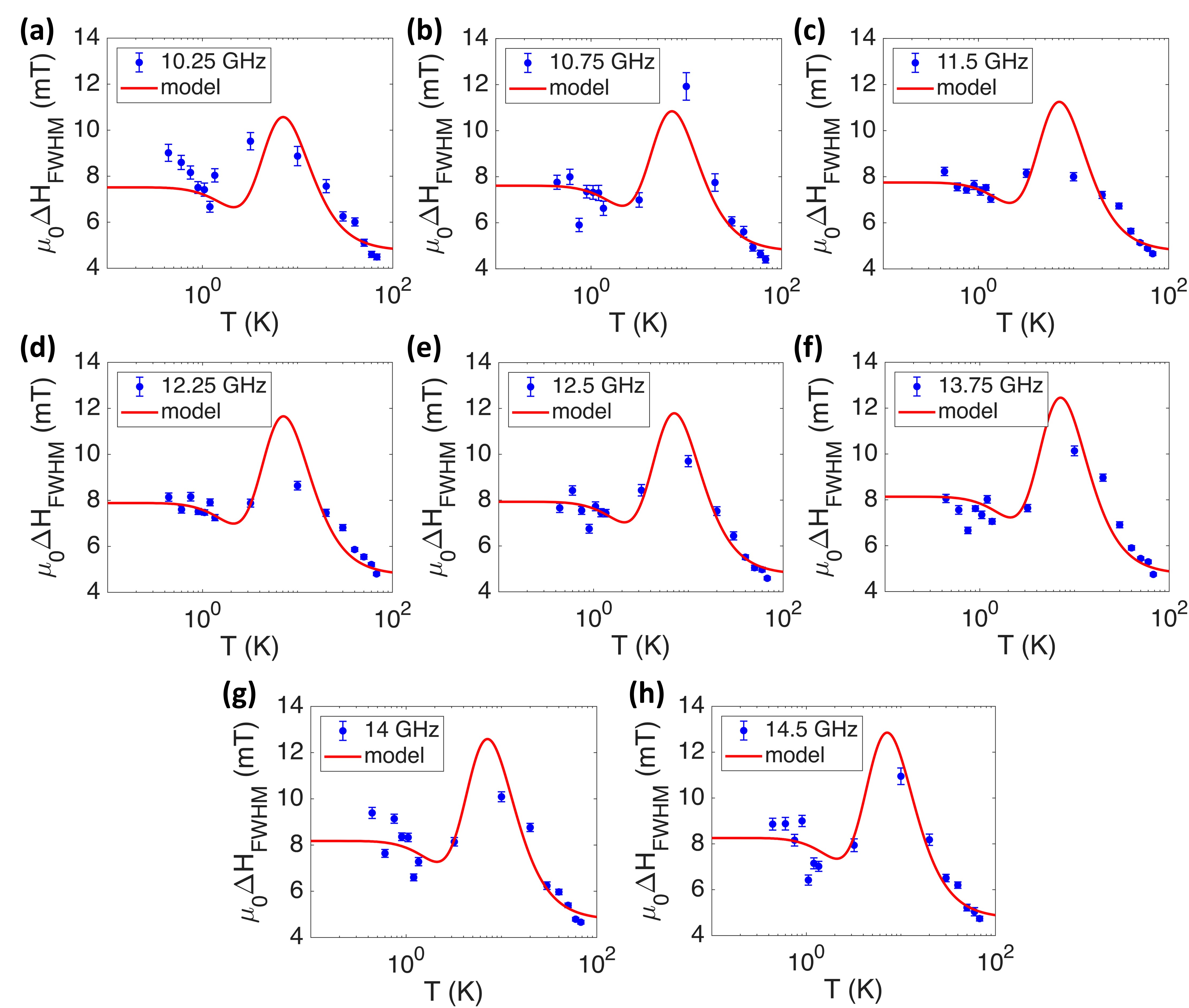}
\caption{(a)-(h) Comparison between the FMR linewidth data at various FMR frequencies and the two--TLS model using a single set of parameters. The TLS exchange splittings are fixed at $\omega_{ex,1} = 26.8~\text{GHz}$ and $\omega_{ex,2} = 296~\text{GHz}$. The  relaxation times at 0 K are the same as those used in Fig.~\ref{fig:fig10}. The parameters $c_1$, $c_2$, and $\eta$ are intrinsic to the material; therefore, we take the mean values of the fitted results across the different FMR frequencies, $\bar{c}_1=1.47$ mT/GHz, $\bar{c}_2=4.75$ mT, $\bar{\eta}=0.88$. }
\label{fig:fig11}
\end{figure}

To obtain a global and self-consistent description, we first identify the pair of exchange splittings $(\omega_{\rm ex,1}, \omega_{\rm ex,2})$ that minimizes the changes of the total sum of squared errors across all measured FMR frequencies. Using this optimized pair, the parameters $(c_1, c_2, \eta)$ are then extracted independently for each individual FMR frequency, as shown in Fig.~\ref{fig:fig10}. The relatively small variance in these extracted parameters demonstrates the robustness of the fitting procedure and the consistency of the optimized exchange splittings.

The mean values $(\bar{c}_1, \bar{c}_2, \bar{\eta})$ together with optimized exchange splittings $(\omega_{\rm ex,1}, \omega_{\rm ex,2})$ provide a single unified parameter set that is consistent with the linewidth data over all FMR frequencies, as illustrated in Fig.~\ref{fig:fig11}.

\subsection{Distribution of exchange splittings for TLS}

The aluminum doping introduces random fluctuations in the chemical environment of the Fe cation impurities; therefore, the exchange splittings of the two types of TLS are actually randomly distributed rather than having fixed values. We therefore decided to compare the fits to the FMR linewidth using a distribution of TLS energy splittings with the fits obtained above using fixed values for the TLS splittings.

\begin{figure}[h!]
\centering
\includegraphics[width=\columnwidth]{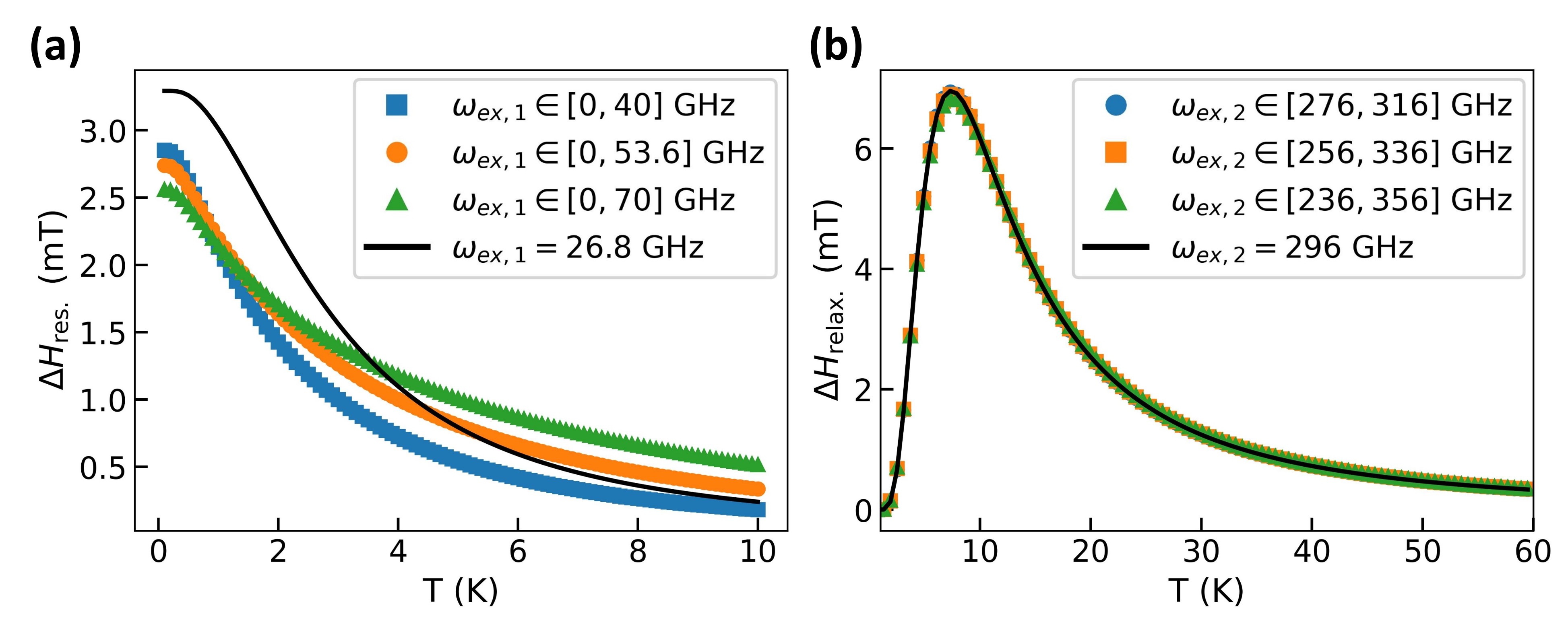}
\caption{TLS-induced FMR linewidth with (a) resonant mode and (b) relaxation mode using fixed (solid black line) and a flat distribution of exchange splitting (colored markers).}
\label{fig:fig12}
\end{figure}

For the case with a distribution of energy splittings, we assume that both types of TLS have a flat distribution of exchange-splitting energies, \textit{i.e.}, $P(\omega) = \text{const}$. The smaller exchange splitting for the resonant mode is assumed to range from 0 to $\omega'_{ex,1}$, while the larger exchange splitting for the relaxation mode is centered at $\bar{\omega}_{ex,2}$ (equal to the exchange splitting used in the previous fitting, $\omega_{ex,2}$) with a finite width $\Delta \omega_{ex,2}$. The FMR linewidths of the resonant and relaxation modes corresponding to flat distributions of exchange splittings are then given by
\begin{align}
\label{q:Appen-F-23}
\Delta H^{ dis.}_{res.} (T ) &=  c_1 \eta^2 \int^{\omega'_{ex,1}}_0 P( \omega) F_1(T;\omega) \ d \omega, \\
\Delta H^{ dis.}_{relax.} (T ) &=  c_1 (1-\eta^4) \int^{\bar{\omega}_{ex,2}+\Delta\omega_{ex,2}}_{\bar{\omega}_{ex,2}-\Delta\omega_{ex,2}} P( \omega) F_2(T;\omega) \ d \omega,
\label{q:Appen-F-24}
\end{align}
where $P(\omega) =1/(\omega_{upper} - \omega_{lower}) $ denotes the normalized flat distribution of exchange splittings with $\omega_{upper},\omega_{lower}$ specifying the upper and lower bounds of the integration.

In Fig.~\ref{fig:fig12}, we compare the TLS-induced FMR linewidth obtained using fixed exchange splittings with that calculated from flatly distributed splittings. For the resonant mode, increasing the upper bound of the distribution does not significantly alter the temperature dependence of the linewidth; however, a smoother decay with increasing temperature is observed for larger upper bounds, as shown in Fig.~\ref{fig:fig12}(a). For the relaxation mode, varying the distribution width $\Delta \omega_{ex,2}$ produces only a marginal change in the resulting linewidth, see Fig.~\ref{fig:fig12}(b). Therefore, we used the fixed--exchange-splitting model for analyzing the experimental data, as it adequately captures the essential features of the temperature profile.

\clearpage
\bibliography{ref}

\end{document}